 \documentclass[preprint,12pt]{elsarticle}
 
 \usepackage{setspace}
\usepackage{amssymb}
\usepackage{amsmath}
\usepackage{color}
\usepackage{tikz}

\usepackage{changes}
\definechangesauthor[name={Maryam Baniasadi}, color={purple}]{MB}

\usetikzlibrary{arrows,shapes,automata,backgrounds,petri,positioning}
\usetikzlibrary{decorations.pathmorphing}
\usetikzlibrary{decorations.shapes}
\usetikzlibrary{decorations.text}
\usetikzlibrary{decorations.fractals}
\usetikzlibrary{decorations.footprints}
\usetikzlibrary{shadows}
\usetikzlibrary{calc}
\usetikzlibrary{spy}
\usetikzlibrary{calc,patterns,decorations.pathmorphing,decorations.markings}
\tikzstyle{spring}=[thick,decorate,decoration={zigzag,pre length=0.3cm,post length=0.3cm,segment length=6}]
\tikzstyle{damper}=[thick,decoration={markings,  
  mark connection node=dmp,
  mark=at position 0.5 with 
  {
    \node (dmp) [thick,inner sep=0pt,transform shape,rotate=-90,minimum width=15pt,minimum height=3pt,draw=none] {};
    \draw [thick] ($(dmp.north east)+(2pt,0)$) -- (dmp.south east) -- (dmp.south west) -- ($(dmp.north west)+(2pt,0)$);
    \draw [thick] ($(dmp.north)+(0,-5pt)$) -- ($(dmp.north)+(0,5pt)$);
  }
}, decorate]
\tikzstyle{process} = [rectangle, rounded corners, minimum width=3cm, minimum height=2.5cm, text 
centered, text width=3cm, draw=black, fill=blue!30]
\tikzstyle{region} = [rectangle, minimum width=3.0cm, minimum height=2.5cm, 
text centered, draw=black, fill=brown!30]
\tikzstyle{arrow} = [thick,->,>=stealth]
\journal{Chemical Engineering Science}

\begin{document}

\begin{frontmatter}



\title{A numerical study on the softening process of iron ore particles in the cohesive zone of an experimental blast furnace using a coupled CFD-DEM method}


\author{Mehdi Baniasadi, Maryam Baniasadi, Gabriele Pozzetti, Bernhard Peters }

\address{University of Luxembourg, Faculty of Science, Technology and Communication, 2, Avenue de l'Université, L-4365 Esch-sur-Alzette, Luxembourg.}

\begin{abstract}
Reduced iron-bearing materials start softening in the cohesive zone of a blast furnace due to the high temperature and the weight of the burden above. 
Softening process causes a reduction of void space between particles. As a result, 
the pressure drop and gas flow change remarkably in this particular zone.
As a consequence, it has a significant influence on the performance of a blast furnace and is needed to be fully characterized. For this reason, the gas rheology along with the deformation of the
particles and the heat transfer between particle-particle and particle-gas should be adequately described. In this paper, the eXtended Discrete Element Method (XDEM), as a CFD-DEM approach 
coupled with the heat transfer, is applied to model complex
gas-solid flow during the softening process of pre-reduced iron ore pellets in an Experimental Blast Furnace (EBF).
The particle deformation, displacement, temperature, and gas pressure drop and flow under conditions relevant to the EBF operations are examined. Moreover, to accurately capture the high gas velocity inlet, 
a dual-grid multi-scale approach is applied. 
The approach and findings are helpful to understand the effect of the softening process on the pressure drop and gas flow in the cohesive zone of the blast furnace.
\end{abstract}

\begin{keyword}
XDEM, CFD-DEM, softening, heat transfer, iron ore pellet, cohesive zone



\end{keyword}

\end{frontmatter}


\section{Introduction} 
\begin{figure}[h]
				\centering
			  \includegraphics[trim = 2mm 12mm 0mm 0mm, clip, width=0.7\linewidth]{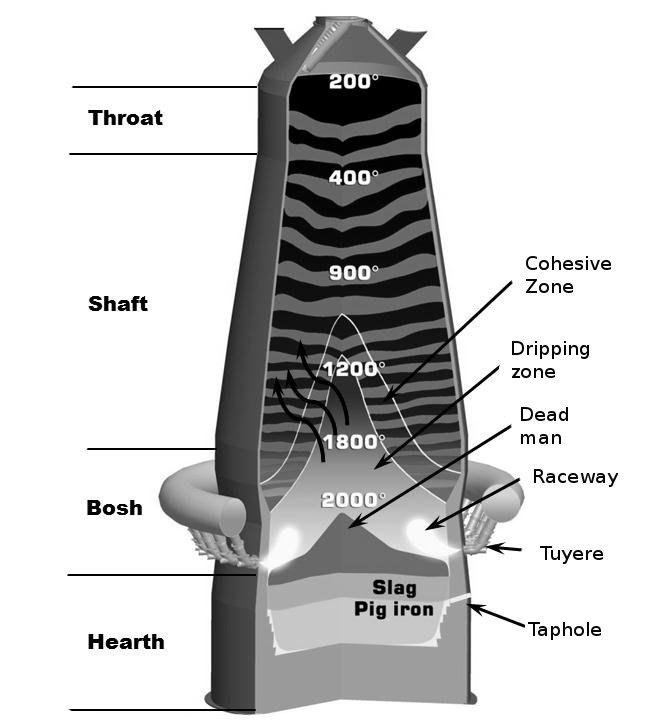}
			  \caption{Sc1hematic diagram of different parts of blast furnace \cite{maryamthesis}.}
			  \label{f:BF1}
			\end{figure}

A blast furnace (BF) is a complex chemical reactor by which iron is produced from iron-bearing materials. 
The BF is charged from the top with alternating layers of iron-bearing materials and coke descending downward by the gravity force. 
A high-velocity stream of preheated air is introduced through the tuyeres in the lower part of the BF to react with the coke.
As a consequence, a counter-current flow of reducing gas, carbon monoxide (CO), that heats up and reduces iron-bearing materials in the shaft, is formed. 
Then, the reduced iron-bearing materials undergo softening and melting in the middle of the BF due to two main reasons: the load of the burden above and the high gas temperature. 
The latter changes the mechanical properties of the particle and the former promotes particles' compression. 
On the other hand, coke particles do not exhibit observable deformation.
The region, where these phenomena take place, is usually referred to as cohesive zone (CZ) as illustrated in Figure~\ref{f:BF1} \cite{maryamthesis}.
As the iron-bearing materials are softened and melted, the local void fraction decreases. This markedly reduces the permeability of the layers of
iron-bearing materials,
affecting the flow distribution and, therefore, the convective heat transfer and temperature distribution.
This complex set of mutual interactions between the gas and the discrete phases determines the main CZ features such as its thickness and position in the BF,
which are known to have a profound effect on the production capability \cite{Aachen}. Therefore,
the description of the characteristics of the softening and melting phenomena is essential for constructing 
a precise blast furnace model. Since it is often not possible to interrupt the BF to investigate details of 
the phenomena occurring inside via purely experimental studies, the numerical simulations can be a practical tool to approach those phenomena. In this contribution, we propose a numerical model that aims to study the softening process.

% In this work, the modeling of the softening process of iron ore pellets is intended. They are one of the main feed sources for blast furnaces.
% The behavior of the partially reduced pellets in the CZ is very complex and depends on a number of variables. 
% The pellets start deforming at a temperature depending on the type of sample, the chemical composition, the reduction degree (RD) and the load. 
% However, Nogueira et al.~\cite{Nogueira2}, Kemppainen et al.~\cite{Kemppainen} and BANIASADI proved that the RD 
% has not a significant effect on 
% the softening of the reduced pellets to RDs between 50-80\%.
% The reduction process in pellets takes place from the surface toward the core of the particle. It leads to the creation of a porous metallic iron shell. 
% The initial oxide melt forms at the core of the pellets. At a certain point, as the volume of melt in the core increases, 
% the shell is cracked and oxide melt will exude from the particle \cite{Bakker, Adema}. At this point, the thickness of pellet layer 
% decreases
% to 50\% of the initial one.
% Due to this complexity few modeling efforts have been carried out. Nevertheless, modeling is an effective way to describe the 
% behavior of the CZ of the BF, which has a harsh condition to be investigated experimentally.

Continuum-continuum modeling has been used to describe the phenomena inside the blast furnace \cite{Yagi, Austin, Austin2}. However, the continuum-continuum method cannot provide 
some significant information such as deformation and temperature of each particle, as well as the local void space between particles which are important, particularly, in the CZ modeling.
As a result, the effects of those on the softening-melting process and the gas flow by those methods are ignored.
Alternatively, Lagrangian-continuum numerical 
models can overcome these shortages. Computational fluid dynamics (CFD) and discrete element method (DEM) as a Lagrangian-continuum method
has been used to simulate the conditions within the raceway or the shaft of the BF~\cite{Miao, Hou, Yu}. Moreover, this approach has been applied to model the whole BF considering 
the phenomena taking place in the CZ with several simplifying assumptions. For instance, Ueda et al.~\cite{Ueda} used the CFD-DEM method to investigate the gas and solid flow in the different cohesive zone shapes by assuming
that the iron ore layers
are impenetrable. Kurosawa et al. \cite{Kurosawa} modeled the BF by considering the softening behaviour of particles due to load within the cohesive zone. 
In \cite{Kurosawa}, the ore particles are allowed overlapping thus reducing the void fraction of the bed and influencing the gas flow due to their softening.
A three-dimensional model was adopted by Yang et al.~\cite{YANG} to
examine the gas-solid flow in a BF with and without the CZ.
However, heat transfer was not included in these works.
Nevertheless, an accurate solution of the temperature distributions for gas and particles are mandatory to predict the softening process correctly and therefore the characteristic of the CZ. 

The deformation of a particle is caused by the change of the the Young's modulus, 
which is strongly dependent on particle temperature. 
As the temperature varies along the blast furnace, the properties of ore particles alter. Therefore, the assumption on the 
constant mechanical properties, particularly, Young's modulus in each part of the BF as was considered in~\cite{Kurosawa} is not realistic. 
Recently, Yang et al.~\cite{Softening_melting, Softening_melting2} implemented temperature dependency of Young's modulus into DEM based on the experimental data carried out by Chew and Zulli~\cite{Chew}. 
The particle deformation, temperature, and gas pressure were examined in the softening and melting of wax balls.
Baniasadi et al. \cite{BANIASADI} investigated the softening process of a type of iron ore pellet experimentally and numerically in a small scale. 
However, the gas rheology was not considered in this study.
To date, there is no research on the simulation of the softening behaviour of the iron-bearing particles along with the gas rheology changes and heat transfer in a large scale. 

As mentioned, the hot gas is introduced from tuyeres and react with coke to produce the reducing gas. This gas flowing upward carries out some vital functions such as to heat up the particles, 
to reduce iron-bearing materials, and to melt 
reduced iron ore particles in different parts of a BF. Therefore, an accurate prediction of 
the fluid dynamic is mandatory to predict not only particle softening but also the whole BF process correctly.
However, obtaining accurate and grid-convergent solutions for the fluid flow within CFD-DEM couplings is a challenging task. 
As shown by Pozetti et al. ~\cite{PozzettiIJMF}, the nature of the volume averaged coupling between CFD and DEM domains can force researchers to use coarse-grids for the fluid solutions 
that do not allow resolving fine-scale fluid structures. This can lead to solutions that cannot ensure grid-convergence nor accuracy.
For the BF modeling, this issue can significantly affect the ability to
reconstruct the high-velocity gas flow~\cite{Miao}. In particular, in~\cite{Miao}, the fundamental role played by the inlet condition on the fluid-dynamic behaviour of the BF was proven; still, 
the numerical experiments were limited to very-low velocities. For this reason, we here adopt 
the dual-grid multiscale approach initially introduced in~\cite{PozzettiIJMF}
that allows us obtaining detailed information on the high-velocity flow field and in particular on its temperature distribution.
This ultimately leads to a more reliable evaluation of the heat transfer between fluid and particles.

In this contribution, the eXtended Discrete Element Method (XDEM) \cite{Peters}, which is a multiscale and multiphysics numerical simulation method is used to model the softening process of iron ore pellets
in the CZ of an EBF. The XDEM is a CFD-DEM method with some extra features such as heat transfer coupling and dual-grid multiscale approach.
The XDEM has been applied to different phenomena taking place in the blast furnace. In particular, it has been validated for 
the heat up and reduction of iron ore pellets in the shaft \cite{reduction, reduction2}, and the flow behavior of molten iron and slag through a packed bed of coke particles in the dripping zone \cite{maryam, maryam2, Maryam4}.
To fulfill our ultimate goal, which is the modeling of a blast furnace using the XDEM, the phenomena inside the CZ also need to be modeled. 
Here, the main focus in this contribution is to model the softening process of iron-bearing materials in a pilot scale.  
Then, the results are analyzed regarding the gas and solid flow, temperature field, particle deformation, and pressure drop.

\label{}

\section{Model description} 
The XDEM method based on a Eulerian-Lagrangian framework is suitable for many engineering applications. 
A comprehensive description of the XDEM platform and its applications can be found in Peters et al. \cite{XDEM-suits}, however, the equations used in this contribution are 
explained here. In the XDEM platform, the solid particles are considered as discrete entities and the gas phase
as a continuous phase that exchange momentum, heat, and mass between each other. In this section, the governing equations for the gas phase are described. Then, the equations to predict the position, 
orientation, and velocity along with temperature distribution of each particle individually are discussed. Later, the correlations used to calculate the coupling terms in 
the XDEM are presented. Finally, the XDEM dual-grid approach is recalled.

\subsection{Fluid Phase Governing Equations}\label{fluidEq}
The governing equations consisting of mass, momentum and energy conservations for the gas phase are presented in Table~\ref{t:continuum}. These equations are derived based on a Eulerian volumetric and time 
averaging method for porous media proposed by Faghri and Zhang \cite{Faghri}.
The void space ($\epsilon$), also known as porosity, refers to the void spaces between the solid particles. 
The importance of this parameter in the porous media is undeniable and its accurate prediction is crucial particularly in the CZ. In the CZ, the iron ore particles are softened, 
which will lead to the creation of low void spaces and as a consequence, this phenomenon can change the gas phase passage locally. The XDEM void space calculation 
method was explained in our previous study \cite{my2}. The coupling parameters such as the fluid-particle interaction forces ($\vec F_d$) and the heat transfer coefficient ($\alpha$) are described in the Table~\ref{table:coupling}.

\begin{table}[ht]
 \caption{The governing equations of the gas phase.}
 \resizebox{1\textwidth}{!}{
 \centering 
 \begin{tabular}{l l}
  \hline
      Void space                & $ \epsilon = \frac{V_g}{V_{cell}} = 1- \frac{V_{s}}{V_{cell}}$\\[1.5ex]
      Conservation of mass      & $ \frac{\partial}{\partial t} (\epsilon \rho_g) + \nabla \cdot (\epsilon \rho_g \vec{v_g} ) = 0  \label{e:continuity} $ \\ [1.5ex]
      Conservation of momentum  & $ \frac{\partial}{\partial t} (\epsilon\rho_g \vec{v_g} ) + \nabla \cdot (\epsilon\rho_g \vec{v_g} \vec{v_g} ) =  $  \\ [1.5ex]
	                        & $ - \epsilon \nabla p + \epsilon\rho_g g + \nabla \cdot [\epsilon \mu (\nabla \vec{v_g} + \nabla (\vec{v_g})^T)] +
	                            - \vec F_d \label{e:momentumg}  	$ \\ [1.5ex]
     Conservation of energy     & $\frac{\partial}{\partial{t}}(\epsilon \rho_g h_g) + \nabla . (\epsilon \rho_g \vec v_g h_g) = $ \\ [1.5ex]
                                & $-\nabla \cdot (\epsilon \lambda_g \nabla T_g) - \alpha (T_g -T_s)   \label{e:energy}$ \\ [1.5ex]
  \hline
 \end{tabular}}
\label{t:continuum}
\end{table}

% ================================POROSITY======================================================================================
% \subsubsection{Porosity calculation} \label{s:porosity}
% The volumetric porosity is estimated by considering the volume of solid particles with their corresponding weight, $\eta_{i,cell}$, in each CFD cell. 
% 
% \begin{equation} %-----porosity-------
% \epsilon_{cell}= 1- \frac{1}{v_{cell}} \sum_{i}^n \eta_{i,cell} V_i
% \end{equation}
% 
% where the weights are calculated based on the method proposed by Xiao and Sun \cite{Xiao2011} and $n$ is the number of particles in each cell. In this method the volume of each particle is weighted
% by its distance from the touching cell centers. The calculated porosity by XDEM in shown in Fig. \ref{f:porosityCalc}. Morever, the average porosity of domain is calcluated by the volumetric integration 
% of local porosity over the cells: 
% 
% \begin{equation}
% \overline{\epsilon} = \frac{1}{v} \int \epsilon_{cell} dv
% \end{equation}
% 
% \begin{figure}[!tbp]
%   \centering
%     \includegraphics[trim = 20mm 40mm 15mm 40mm, clip, angle=0, width=0.9\textwidth]{Figures/porosityCalc.pdf}
%     \caption{The volumetric average porosity calculated by the XDEM \cite{my2}.}
%     \label{f:porosityCalc}
%   \end{figure}
%================================================================================================================================

\subsection{Particles Governing Equations} 
The motion and position of each particle are predicted using a discrete element method based on the soft-sphere model, 
in which the particles are allowed to be overlapped in order to represent 
the softening behaviour of pellets. In this method, Newton's 
and Euler's second law for translation and rotation of each particle, which are given in Table~\ref{table:DEM} are integrated over time.
The contact force $\vec{F}_{i}^c$ of a particle is described by the Hertz-Mindlin model \cite{hertz, mindlin}, which is a non-linear model for the contact force. 
The $\vec{F}_{i}^c$ is the sum of all normal and tangential collision forces generated while colliding with the neighboring particles. This model was successfully employed by 
Baniasadi et al.~\cite{BANIASADI} and Yang et al. \cite{Softening_melting, Softening_melting2}. 
Based on this model, the contact force is non-linearly proportional to the overlap between particles and the mathematical model to calculate the contact force is also given in Table~\ref{table:DEM}.

% 
% The XDEM predicts both dynamics and thermodynamics of a particulate system. The thermodynamic state of a particle may simply include an 
% internal temperature distribution, but may also contain transport of species due to diffusion or convection in a porous matrix in conjunction
% with thermo-chemical conversion due to reaction mechanisms. Nevertheless, here just the required equations for softening process is mentioned. 
% The velocity, position, and acceleration are calculated in the dynamics module while the temperature in the conversion module of the XDEM.

\begin{table}[ht]
 \caption{The equations in the dynamic module of the XDEM platform.}
 \resizebox{0.9\textwidth}{!}{
 \centering 
 \begin{tabular}{l l}
  \hline
   Newton's second law of translation   & $m_i \frac{d \vec{v_i}}{d t} = m_i \frac{d^2 \vec{x_i}}{dt^2} = \vec{F}_{i}^c + \vec{F}_{i}^g + \vec{F}_{i}^d$ \\ [1.5ex]
   Euler's second law of rotation   & $I_i \frac{d \vec{\omega}_i}{d t} = \sum_{j=1}^n \vec{M}_{i,j}$ \\ [1.5ex]
   Normal contact force   & $\vec{F}_{i,j}^{c,n} = -k_n \delta_{n}^{\frac{3}{2}} - c_n \delta_{n}^{\frac{1}{4}} \dot{\delta}_{n}$ \\ [1.5ex]
   Overlap  & $\delta_{n} = r_i + r_j - |\vec{x}_{i} -\vec{x}_{j}|$ \\ [1.5ex]
   Normal spring stiffness  & $k_n = -\frac{4}{3}E_{i,j}\sqrt{R_{i,j}}$ \\ [1.5ex]
   Normal viscous damping coefficient  & $c_n = ln e \sqrt{\frac{5m_{i,j}k_n}{\pi + ln e^2}}$  \\ [1.5ex]
   Effective radius and reduced mass  &  $\xi_{i,j} = \frac{\xi_1 \xi_2}{\xi_1 + \xi_2} \quad, \quad \xi = R,m$  \\ [1.5ex]
   Effective Young's modulus & $\frac{1}{E_{i,j}} = \frac{1-{v_i}^2}{E_i} + \frac{1-{v_j}^2}{E_j}$ \\ [1.5ex]
   Tangential contact force  & $\vec{F}_{i,j}^{c,t} = min [k_t \delta_{t} + c_t \dot{\delta}_{t} \quad, \quad \mu \vec{F}_{i,j}^{c,n}]$ \\ [1.5ex]
   Tangential stiffness &  $k_t = -8 G_{i,j}\sqrt{R_{i,j} \delta_{n}}$ \\ [1.5ex]
   Tangential dissipation coefficient & $c_t = ln e \sqrt{\frac{5 (4 m_{i,j} k_t)}{6 (\pi + ln e^2)}}$ \\ [1.5ex]
   Effective shear modulus &  $\frac{1}{G_{i,j}} = \frac{2-v_i}{G_i} + \frac{2-v_j}{G_j}$ \\ [1.5ex] 
  \hline
 \end{tabular}}
\label{table:DEM}
\end{table}

% \begin{figure}[!tbp]
%   \centering
%     \includegraphics[trim = 30mm 160mm 0mm 10mm, clip, angle=0, width=1.05\textwidth]{Figures/collision.pdf}
%     \caption{Schematic illustration of collision for two spherical particles \cite{my2}.}
%     \label{f:collision}
%   \end{figure}
% Figure \ref{f:collision} shows two particles $i$ and $j$ with position vectors $\vec{x}_{i}$, $\vec{x}_{j}$,
% and radii $r_i$ and $r_j$ are in colliding condition.

\begin{figure}[!tbp]
  \centering
    \includegraphics[trim = 0mm 2mm 0mm 2mm, clip, angle=0, width=7cm]{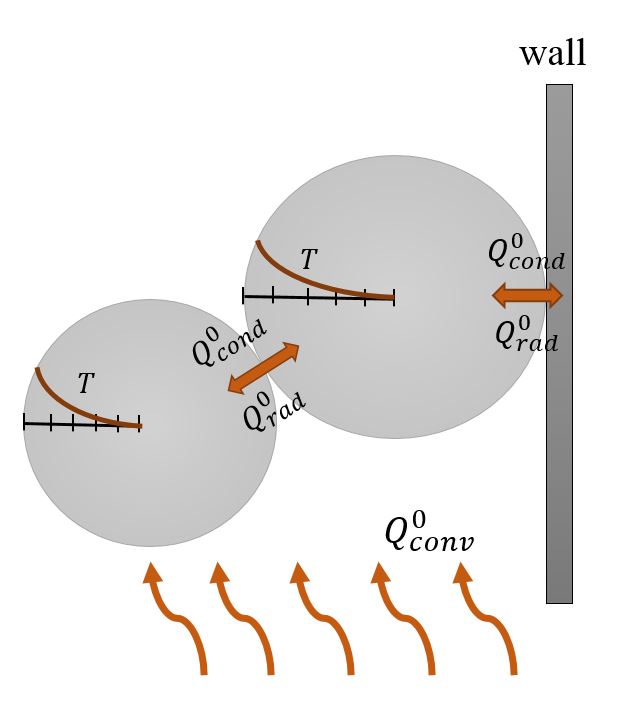}
    \caption{Illustration of heat exchanges between particles and wall.}
    \label{f:heat}
\end{figure}

The particle softening behaviour is closely related to the temperature. Therefore, the key step to applying CFD-DEM to study the 
phenomenon in the CZ is that heat transfer must be considered. The XDEM has been already enhanced and validated for fluid-particle heat up process \cite{reduction, reduction2, my, shrinking, mahmoudiDrying}. 
To investigate the softening process in high temperature, the heat exchange between particle-particle, and particle-gas along with the temperature gradient
within the particle should be taken into account as illustrated schematically in Figure \ref{f:heat}. The conversion module of the XDEM provides those by considering 
a system of one-dimensional and transient energy equation and the boundary conditions as written in Table~\ref{table:coupling}.
$q_{cond}'$ and $q_{rad}'$ account for conductive heat and radiation transport through  
physical contact with the wall and/or particles. $F_{P,j}$ is the view factor between the particle p and its neighbor j. For a 
more detailed derivation of the above equations the reader is referred to Peters \cite{Peters}.
\begin{table}[ht]
 \caption{The equations in the conversion module of the XDEM platform.}
 \resizebox{0.8\textwidth}{!}{
 \centering 
 \begin{tabular}{l l}
  \hline  
   Energy equation &  $\frac{\partial}{\partial t} \left( \rho h  \right) = \frac{1}{r^n}\frac{\partial}{\partial r}\left( r^n \lambda \frac{\partial T}{\partial r} \right) \label{eq:energyyy}$ \\ [1.5ex] 
   Boundary condition    &  $\lambda \frac{\partial T}{\partial r} |_{r = 0} = 0 $ \\ [1.5ex] 
                         &  $ \lambda \frac{\partial T}{\partial r} |_{r = R} = \alpha (T_{inf} -T_s) + q_{cond}' + q_{rad}$ \\ [1.5ex]
                         &  $ q_{cond}' = \frac{\frac{T_p - T_j}{\Delta x_{pj}}}{1/\lambda_p + 1/\lambda_j}$ \\ [1.5ex] 
                         &  $ q_{rad}' = \sum_{j}^{M} F_{P,j} ({T_P}^4 - {T_j}^4)$ \\ [1.5ex] 
 \hline
 \end{tabular}}
\label{table:conversion}
\end{table}

\subsection{Fluid-particle Heat and Momentum transfer Models} \label{s:fluid-particle}

In order to calculate the convective heat transfer between a sphere and fluid flow, the classical Nu-relation, Ranz and Marshall \cite {Ranz} is used in this study. 
The other coupling term is the momentum interaction term, which is the drag force, $\vec{F_d}$, depends on the relative velocity of the solid particle and fluid
and the effect of presence of neighbouring particles. The drag force coefficient, $\beta$, is calculated using modified 
Ergun correlation \cite{ergun}. The Ergun equation is a popular eqaution in gas-solid flow studies to describe gas pressure drop through the packed bed of 
particles and has been also used for the 
softening process by Kurosawa et al.\cite{Kurosawa}. The models are described in Table~\ref{table:coupling}.

\begin{table}[ht]
 \caption{The coupling model of the XDEM platform.}
 \resizebox{0.8\textwidth}{!}{
 \centering 
 \begin{tabular}{l l}
  \hline  
   Heat transfer coefficient            &  $\alpha = \frac{Nu \lambda}{2r_p}$ \label{eq:alfa} \\ [1.5ex] 
   Ranz-Marshall correlation \cite{Ranz}&  $ Nu = 2 + 0.6 Pr^{\frac{1}{3}}Re^{\frac{1}{2}}$ \\ [1.5ex]
   Drag force                           &  $\vec{F_d} = \frac{\beta V_p}{(1-\epsilon)}(v_g - v_p) \label{eq:43}$  \\ [1.5ex] 
%    Ergun correlation \cite{ergun}       &  $ \beta = 150\frac{({1-\epsilon})^{2}}{\epsilon}\frac{\mu_{g}}{(d_p) ^ {2}}$  \\ [1.5ex] 
%                                         & $+1.75(1-\epsilon)\frac{\rho_{g}}{d_{p}}\mid v_g - v_p \mid $ \\ [1.5ex] 
  \hline
 \end{tabular}}
\label{table:coupling}
\end{table}

% 
% \begin{equation}\label{eq:44}
%     \beta = 
% 150\frac{({1-\epsilon})^{2}}{\epsilon}\frac{\mu_{g}}{(d_p) ^ {2}}+1.75(1-\epsilon)\frac{\rho_{g}}{d_{p}}\mid v_g - v_p \mid 
% \end{equation}
% 
% where $C_d$ is given by,
% 
% \begin{equation}\label{eq:45}
%     C_d= 
%     \begin{cases}
%     \frac{24}{Re} [1+0.15 (Re)^{0.687}], & \text{if } Re\textless 1000\\
%     0.44,              & \text{if } Re \textgreater 1000\\ 
%     \end{cases}
% \end{equation}
% 
% and the particle Reynolds number is,
% 
% \begin{equation}\label{eq:46}
%   Re = \frac{\epsilon \rho_g \mid v_g - v_p \mid d_p}{\mu_g}
% \end{equation}

\subsection{Dual-grid multiscale CFD-DEM coupling} \label{dualMesh}

\begin{figure}[ht!]
\centering
\begin{tikzpicture}[node distance=2cm]
%%%%%%%%%%%%%%coarseMESH and description%%%%%%%%%%%
 \node (coarsemesh) [region] {Coarse Grid};
 \node (descrCoarse) [process, right of=coarsemesh, xshift=1.65cm] 
{Calculating continuum fields from Lagrangian entities.\\ -Solving fluid-particles 
coupling};
 \node (mapping) [ultra thick, draw=blue, ellipse, minimum width=10pt,
    align=center, right of=coarsemesh, xshift=1.65cm, yshift=-3cm] 
{Grid-to-Grid\\ Interpolation};
%%%%%%%%%%%%%%%%%%%FINEMESH
 \node (fineMesh) [region , below of=coarsemesh, yshift=-3.9cm] {Fine Grid};
 \node (descrFine) [process, right of=fineMesh, xshift=1.5cm] {- Mass \\
 - Momentum\\- Energy \\
 conservation};
% %%%%%%%%%%%%%ARROWS%%%%%%%%%%%%%%%%%%%%%%%%%%%%%%%%%
 \draw[-latex] (coarsemesh) to[bend right=10] node[above,rotate=90] {Particle 
fields} (fineMesh);
 \draw[-latex] (fineMesh) to[bend right=10] node[below,rotate=90] {Fluid 
Solution} (coarsemesh);
\end{tikzpicture}
 \caption{\label{MULTISCALEWORKFLOW} Diagram of the solution procedure within a dual-grid
 multiscale approach. The two boxes on the left represent the grids associated the bulk
 and the fluid-fine scale, while the 
arrows show schematically the communication between the 
scales.
The coarse grid (top) is used to map the Lagrangian field of the DEM into an 
Eulerian reference and to solve the fluid-particle interaction.
Particle-related fields are mapped to the supporting domain (bottom) where 
a finer grid is used to solve the fluid equations.}
\end{figure}
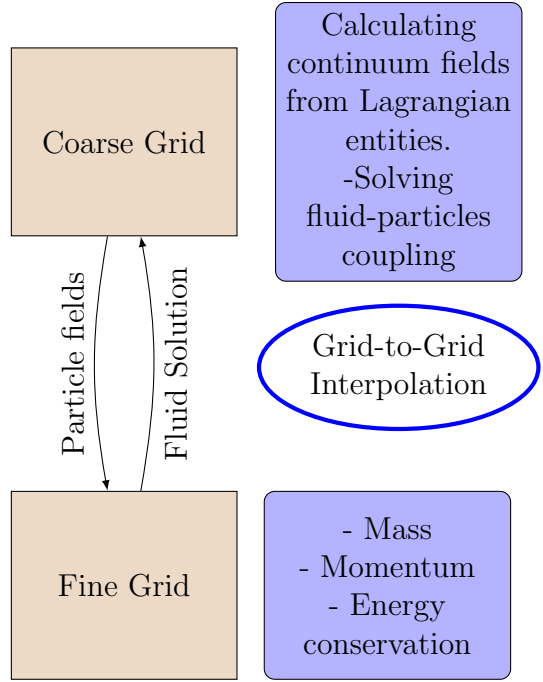

In a dual-grid multiscale CFD-DEM coupling as introduced in~\cite{PozzettiIJMF}, two length-scales are identified within the particle-laden flow.
Fluid-particle interactions are resolved on a bulk scale that is the one typically used for CFD-DEM couplings based on the volume averaging technique.
The main assumption on this lengthscale consist in considering particles as zero-dimensional entities interacting with the fluid with local exchanges
of mass, momentum and energy. In order to enforce this hypothesis, a coarse and uniform grid is the most approapriate choice.
The fluid equations are discretized on fluid fine-scale that is different from the previous one and chosen as the scale at which 
grid-convergent results can be ensured. For this, fine grid discretizations are normally required, particularly for configurations 
featuring high-velocity flows and complex multiphase phenomena~\cite{PozzettiIJMF}.

As shown in Figure~\ref{MULTISCALEWORKFLOW}, a coarse and uniform grid is adopted to resolve the bulk scale, namely the calculation 
of the void space field, drag, and heat source according to what presented in sections~\ref{fluidEq} and \ref{s:fluid-particle}. %~\ref{s:porosity}, \ref{s:fluid-particle}, and~\ref{s:convective_heat_transfer}.  
A fine grid is instead used to solve the fluid flow equations at the fine-scale, namely the conservation of mass, momentum and energy as proposed in section~\ref{fluidEq}.
An interpolation strategy ensures the communication between the two grids
i.e. the two lengthscales.
As originally done in~\cite{PozzettiIJMF}, the interpolation between meshes is performed via the \textit{OpenFOAM-extend} $meshToMesh$ library.

This multiscale approach to CFD-DEM couplings was originally proposed in~\cite{PozzettiIJMF,PozzettiICNAAM2016} and was succesfully applied to study 
different complex flow configurations~\cite{bpadditivemanufacturing, PozzettiPowderTec}. In this contribution, the dual-grid multiscale approach is chosen
in order to obtain grid-independent solution for the gas phase flow within the BF. As shown in~\ref{GSInteraction}, grid-independent solutions are only obtained
with discretization finer than the single-scale one.

\section{Results and Discussion}
In this work, the modelling of the softening process of pre-reduced iron ore pellets in the CZ is intended. Pellets are one of the primary feed sources for blast furnaces.
It is believed that pellets are reduced by the reducing gas in the shaft, above the CZ, to a Reduction Degree
(RD) more than 50\% and less than 80\% \cite{Bakker, KemppainenThesis}. The RD is defined as the ratio of the oxygen removal and the total mass of removable oxygen initially present in the iron ore.
Then, they are softened and molten and generating two different liquids: molten iron and primary slag. The behaviour of the partially reduced pellets during the softening process might depend on several variables. 
The pellets start deforming at a temperature that depends on the type of sample, the reduction degree (RD) and the load~\cite{Adema}. 
Nonetheless, Baniasadi et al. \cite {BANIASADI}, Nogueira et al.~\cite{Nogueira2}, and Kemppainen et al.~\cite{Kemppainen} proved that the RD 
has not a significant effect on 
the softening of the reduced pellets to RDs between 50-80\%. Moreover, the best estimate for the load, the weight of the burden above the CZ, is $100kPa$ \cite{Bakker, KemppainenThesis}.
Therefore, the type of the pellet and its composition are the main factors in the softening process. In this study, we selected a type of pellets which is an acid pellet
from a working plant. The composition of the pellet is shown in Table \ref{table:composition}. It is assumed that the pellets are partially reduced to the RD of 70\%.The composition
of the reduced pellet is obtained according to the iron oxide reduction reactions~\cite{reduction, reduction2} by using a simple mass balance calculation as shown in Table \ref{table:composition_reduced}. 
It is assumed the slag consists of $FeO$, $SiO_2$, $CaO$, $MgO$, and $Al_2O_3$.

\begin{figure}[!tbp]
  \centering
    \includegraphics[trim = 0mm 0mm 0mm 0mm, clip, angle=0, width=14cm]{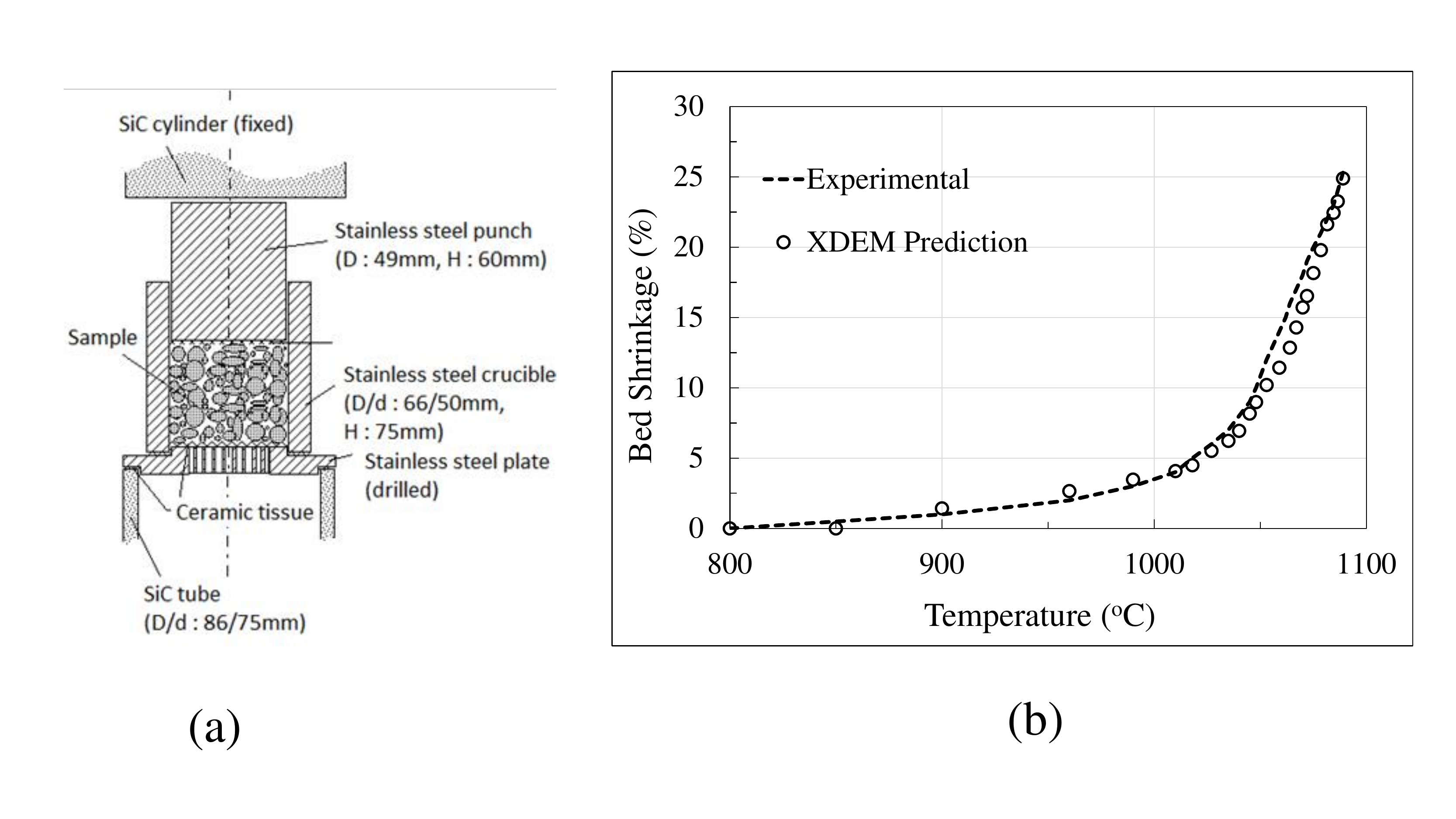}
    \caption{(a) The experimental setup. (b) Comparison of the XDEM results and experimental data for the bed shrinkage.}
    \label{f:exp_simulation}
\end{figure}

To model the softening process of pellets in a real BF, the XDEM should be first validated for this process. 
% To model the softening process of pellets at high temperature, appropriate mechanical properties for the studied pellets, particularly, Young's modulus is a necessity.
For this reason, the softening experiments for a lab-scale packed bed of reduced pellet were carried out. 
A small crucible filled by the pellet particles as illustrated in Figure~\ref{f:exp_simulation} (a) 
was placed inside a big furnace. A stainless steel punch was used to exert $100KPa$ load on the top of particles. 
The temperature of the furnace is increased gradually from $800$ to $1100^{o}C$, and the hight changes of the bed of particles are recorded over temperature. 
The process was simulated by the XDEM and the results for the evaluation of the bed
shrinkage over temperature was compared with the experimental results as shown in Figure~\ref{f:exp_simulation} (b). Consequently, an appropriate relationship between Young's modulus and temperature for the studied pellets has been established. 
% Pellets were screened in the range of $10-12 mm$ sizes.
For more details of the experiments and simulation results, the reader is referred to \cite {BANIASADI}.
In this study, the main target is to apply the validated model for describing the softening process in a larger scale close to a real BF condition.

\begin{table}[ht]
 \caption{Chemical composition of the iron ore pellet.}
 \resizebox{0.25\textwidth}{!}{
  \centering 
 \begin{tabular}{l l}
  \hline
   Component&  Wt(\%)  \\ \hline
    $Fe_2O_3$ & 92.43  \\ [1ex]
     $FeO$  & 0.57  \\ [1ex]
      $SiO_2$  & 4.22  \\ [1ex]
      $CaO$  & 1.42  \\ [1ex]
      $MgO$  & 0.55  \\ [1ex]
            $Al_2O_3$  & 0.39  \\ [1ex]
       $inerts$  & $rest$  \\ [1ex]
   
  \hline
 \end{tabular}}
\label{table:composition}
\end{table}

\begin{table}[ht]
 \caption{Chemical composition of the reduced iron ore pellet.}
 \resizebox{0.25\textwidth}{!}{
  \centering 
 \begin{tabular}{l l}
  \hline
   Component&  Wt(\%)  \\ \hline
    $Fe$ & 56.1  \\ [1ex]
     $Slag$  & 43.9  \\ [1ex]
  
  \hline
 \end{tabular}}
\label{table:composition_reduced}
\end{table}

% The temperature dependence of Young's modulus was chosen by extrapolation of our small-scale experiment results.
% It is generally accepted that the deformation rate of iron ore layers changes linearly. 

\subsection{Simulation results}

\subsubsection{The simulation setup}

% Typically, it takes 1.6 tonnes of iron ore and around 450kg of coke to produce a tonne of pig iron, the raw iron that comes out of a blast furnace. 
% size of layers is roughly equal.
\begin{figure}[!tbp]
  \centering
    \includegraphics[trim = 0mm 0mm 0mm 0mm, clip, angle=0, width=12cm]{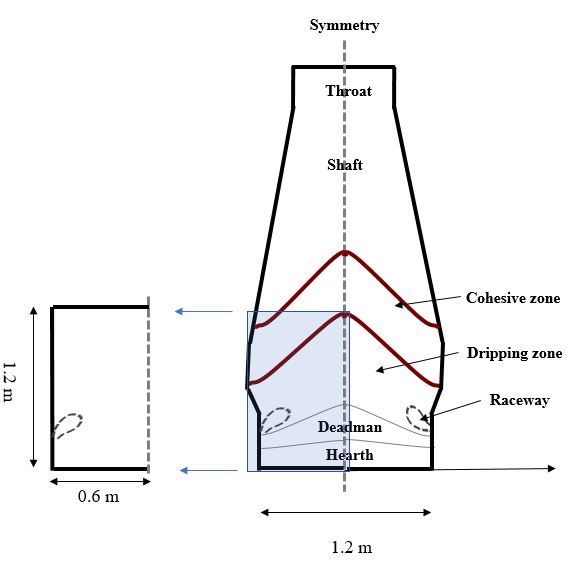}
    \caption{Schematic diagram of different parts of the LKAB's EBF and the zone defined to investigate the gas-solid interaction during the softening process.}
    \label{f:BF}
\end{figure}

\begin{figure}[!tbp]
  \centering
    \includegraphics[trim = 60mm 0mm 0mm 0mm, clip, angle=0, width=18cm]{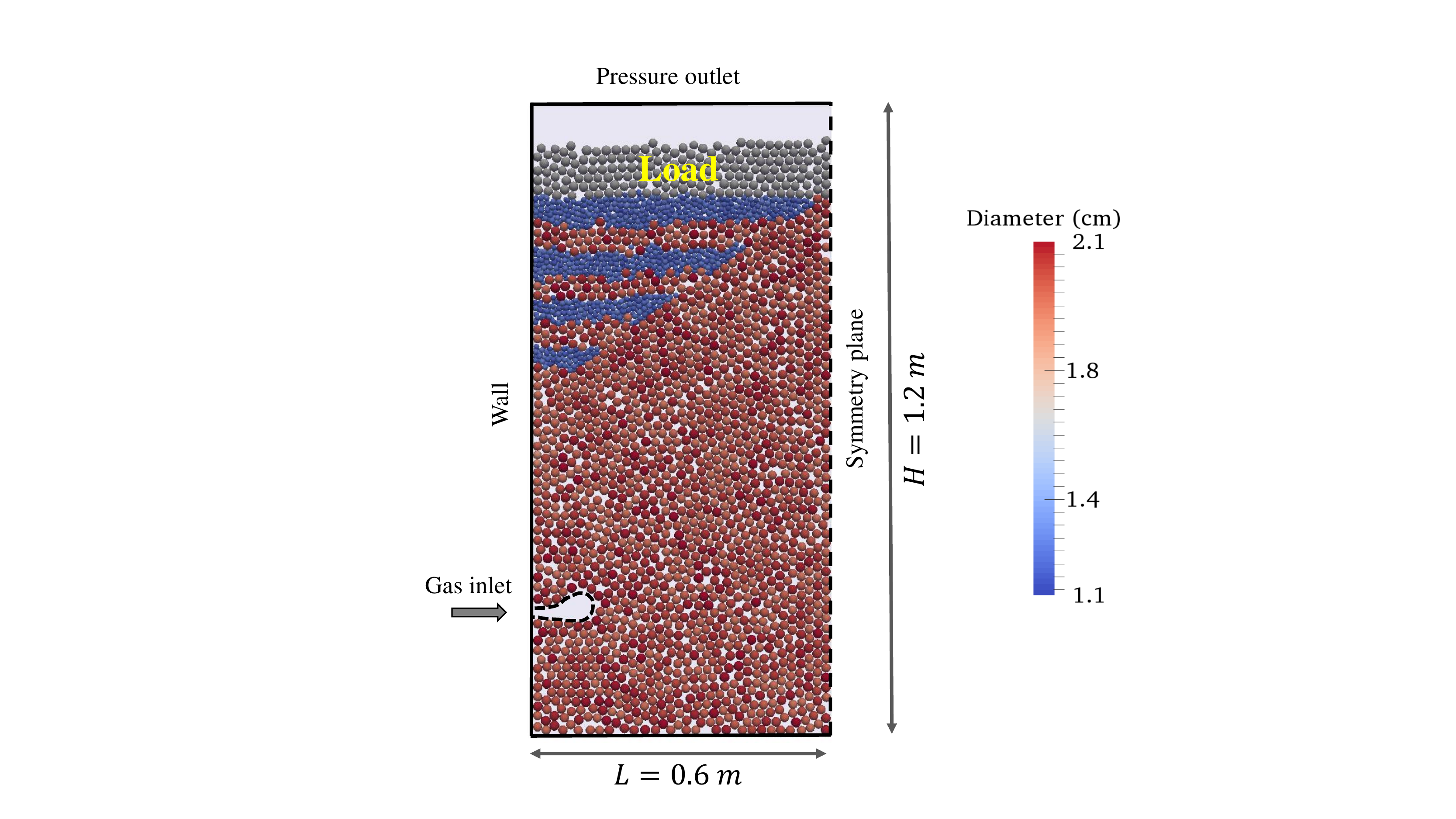}
    \caption{The model geometry filled with the alternating layers of ore and coke above the CZ and only coke below that. The top layer acts as a load representing the weight of burden above the CZ.}
    \label{f:particlePositions}
\end{figure}

To verify the gas-solid interaction during the softening process of the iron ore pellets in a larger scale, the bottom part of the LKAB's Experimental Blast Furnace (EBF) \cite{EBF, EBF1} is considered.
LKAB's EBF was built primarily for product development of iron ore pellets in 1999. 
It is used for the evaluation of the burden
materials as well as the investigation of new blast furnace operational concepts and
equipments \cite{EBF2}.
% Likely a real BF, EBF  is also a cylinder whose diameter varies by its hight. 
The EBF has a working volume of $8.2 m^3$ with the diameter of $1.2m$ at tuyere level. 
% A bell-less top is used for material charging to the EBF. 
The operation of the EBF is similar to a commercial blast furnace, although with a smaller size.
A two-dimensional schematic of the EBF is shown in the right hand side of Figure~\ref{f:BF}. The part of the EBF considered in our simulation is illustrated in the left hand side of Figure~\ref{f:BF}, 
while neglecting the radial variations.
% It produces approximatively 35 tons of hot metal/day.
To reduce the computational time, a symmetric 2D model is considered which is often used in the BF modelling \cite{Zhou1, Zhou2, YangThesis}. 
This section was provided for the simulation with the alternating layers of ore and coke above the CZ and only stationary cokes below that.
This can be performed by the dynamic module of the XDEM, 
which tracks the motion of each particle under gravity and its dynamic collisions with other particles and boundaries.
Figure~\ref{f:particlePositions} represents the setup used in this simulation. The coke and pellet are coloured by their sizes. The pellets' size range form $11mm$ to $13mm$ and 
the cokes $19$ to $21mm$~\cite{EBF}.
Cokes in front of the tuyere are assumed being consumed by combustion thus creating voidage. 
% In this process, the oxygen in the blast furnace is transformed into gaseous carbon monoxide. The
% The resulting gas has a high flame temperature between $2400K$ and $2600K$. 
The shape of the CZ is dependent on the burden distribution. Two basic types of the CZ can be discriminated: inverted V and W \cite{BF_book}.
In this study, inverted V shape is assumed for the CZ. In this type, the center of the furnace only contains coke. Therefore, in the center of the furnace, no softening and melting would occur.

\subsubsection{Gas-solid interaction characterization} \label{GSInteraction}
The generated setup is provided by specifying an inlet gas at a temperature of $2500K$ and at a velocity of $150 m/s$. The initial temperature for 
the coke and pellet particles and the gas is considered to be $1250K$ because 
below that the softening of the pellets is negligible according to the results of the small-scale softening experiments \cite{BANIASADI}. 
Table~\ref{table:properties} shows the required parameters used in the simulation. The simulation time was $40$ seconds when the temperature of pellets in the upper layer reaches 
to $1377K$, which is the melt onset temperature of the studied pellet estimated in the previous study \cite{BANIASADI}.
The time is shown as a dimensionless form as $t^*$. It should be noted that, the liquid phases are not considered in this study. 

\begin{figure}[!tbp]
  \centering
    \includegraphics[trim = 15mm 80mm 0mm 80mm, clip, angle=0, width=15cm]{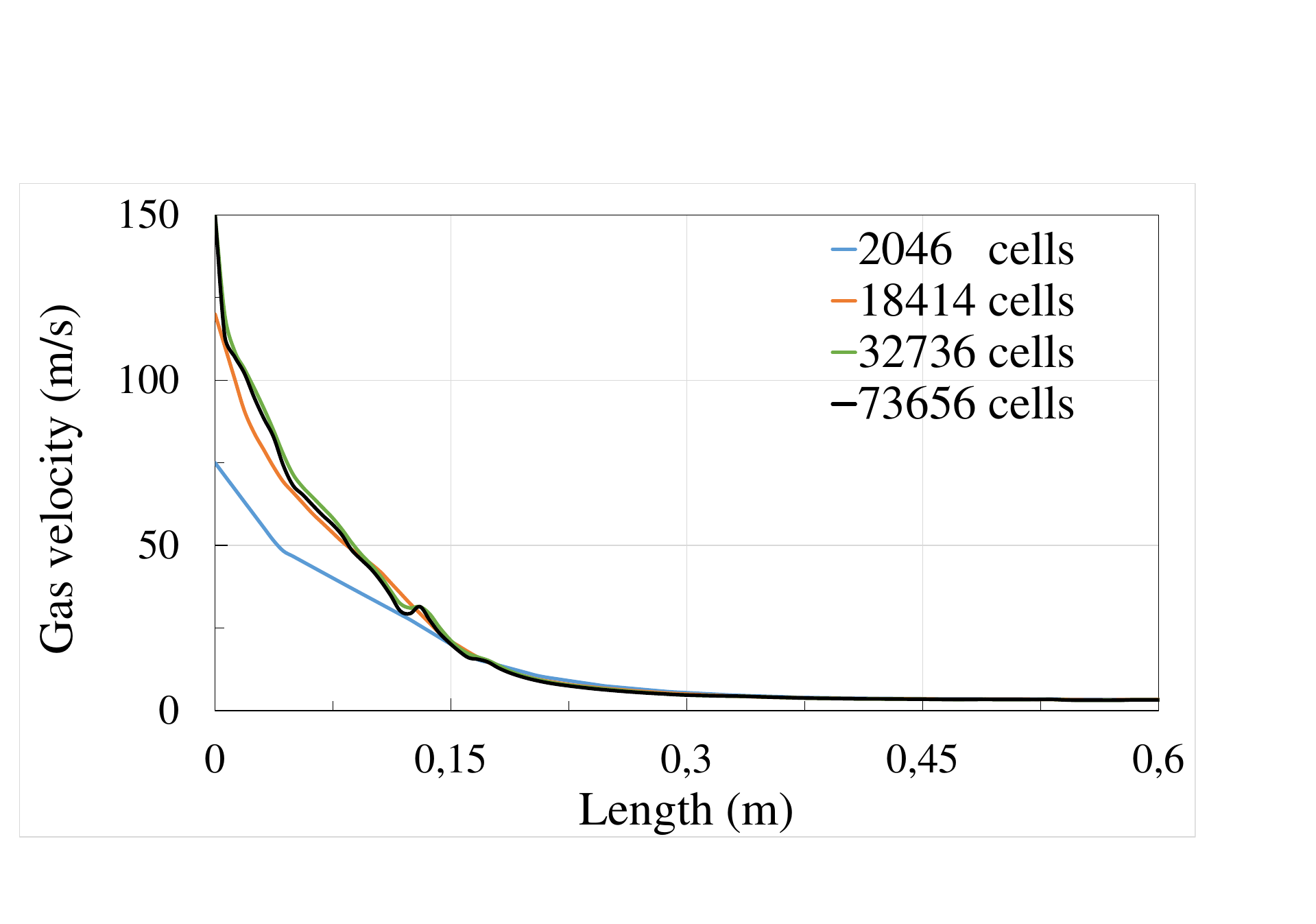}
    \caption{The gas velocity magnitude for four different cell numbers over the length at height of $0.25 m$ (tuyere level) and $t^*=1 $.}
    \label{f:VelocityCheck}
\end{figure}

\begin{figure}[!tbp]
  \centering
    \includegraphics[trim = 15mm 0mm 0mm 0mm, clip, angle=0, width=15cm]{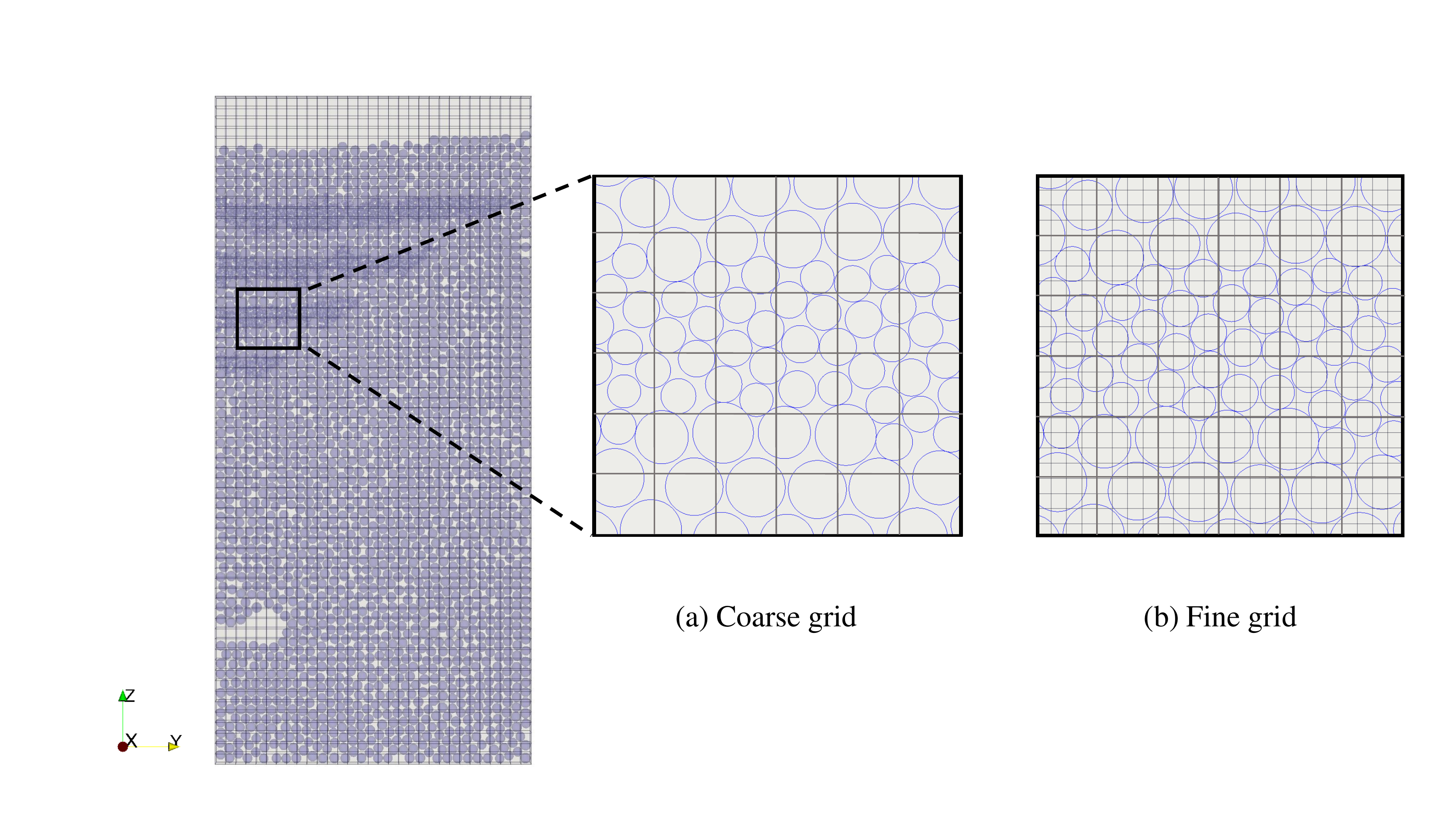}
    \caption{Representation of the number of cells in the (a) coarse grid with $2046$ cells and (b) fine grid with $32736$ cells.}
    \label{f:Coarsefine}
\end{figure}

The velocity of the injected gas from tuyeres of the LKAB's EBF is high.
Thus, the appropriate reconstruction of fluid dynamics requires refining the CFD grid.
Moreover, the calculation accuracy of gas velocity and the temperature is substantial. These variables influence the convective heat transfer between particle and gas which is responsible
for the heat up, reduction and melting of particles.
Therefore, it may change the softening and melting behaviour of iron-bearing materials and as a result, the BF productivity. To achieve this high level of accuracy, the scheme explained in 
section~\ref{dualMesh} is used.
To find the best CFD construction, the grid independence of the solution was studied on the gas velocity at the tuyere level as shown in Figure~\ref{f:VelocityCheck}.
The number of cells was varied in the range of 2046--73656. An underestimation of the gas velocity in the cases with coarser meshes is observed. 
The simulation was found to be grid independent when the number of cells exceeded 32736 where no significant changes were observed by refining the mesh. 
% This is also shown by the gas velocity magnitude contours for the whole domain presented in Figure~\ref{f:grid_check_whole}. 
% Gradient changes in the inlet and the softening areas can be observed.
Therefore, the grid system having 16 times cells more than the coarse mesh was used to resolve the details of flow, and 
temperature fields. The model geometry with coarse and fine meshes is represented in Figure~\ref{f:Coarsefine}.

% \begin{figure}[!tbp]
%   \centering
%     \includegraphics[trim = 0mm 0mm 0mm 0mm, clip, angle=0, width=16cm]{Figures/FinalFigs/grid_check_whole.pdf}
%     \caption{The gas velocity magnitude contours for four different cell numbers at $t^*=1 $. }
%     \label{f:grid_check_whole}
% \end{figure}

\begin{figure}[!tbp]
  \centering
    \includegraphics[trim = 5mm 0mm 0mm 0mm, clip, angle=0, width=17cm]{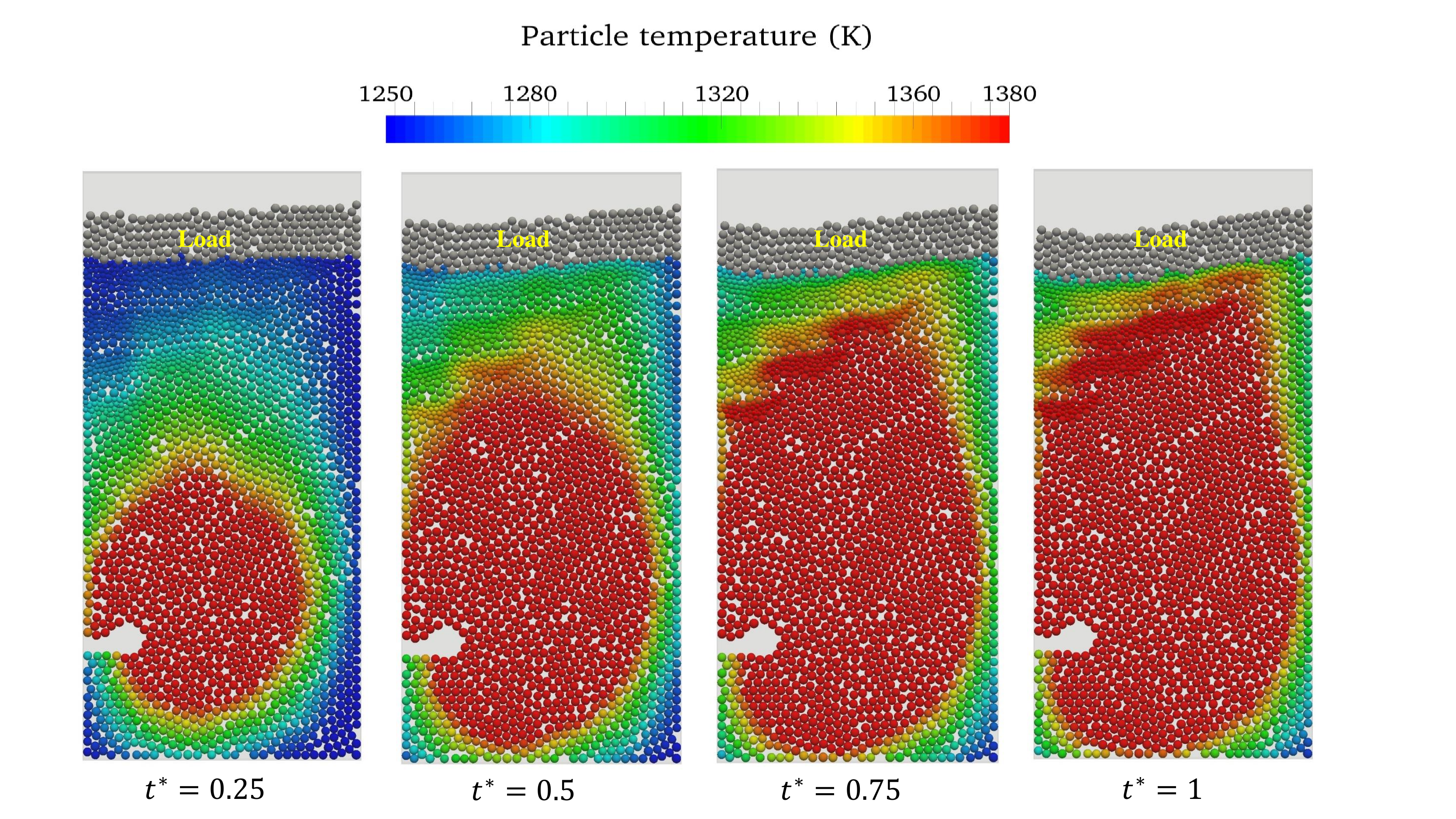}
    \caption{Particle temperature evaluation over time.}
    \label{f:particle_temperature}
\end{figure}

The most significant phenomena experienced by the
softening are the heat transfer and the gradient changes of permeability, pressure drop and gas flow path, which can be attributed to 
the deformation of particles.
Therefore, temperature and micro-dynamics of particles like displacement and deformation cause voidage reduction, thus affect the gas flow field, 
and pressure drop. These phenomena are used to demonstrate the general features of the softening process. To describe them accurately, a validated and verified approach for particle heating up 
by a hot fluid, and the rheology of gas facing particles is needed.
The XDEM which is a CFD-DEM method coupled with the heat transfer has been validated for the heat transfer in packed beds including convective, particle-particle conduction, and particle radiation 
\cite{BANIASADI, reduction, reduction2}.
Besides, 
the pressure drop of a gas flow in a packed bed for different particle sizes was appropriately described by Baniasadi and Peters~\cite{maryam2}.
Therefore, the XDEM is a reliable
tool to be applied to the softening process.

\begin{figure}[!tbp]
  \centering
    \includegraphics[trim = 0mm 30mm 0mm 0mm, clip, angle=0, width=18cm]{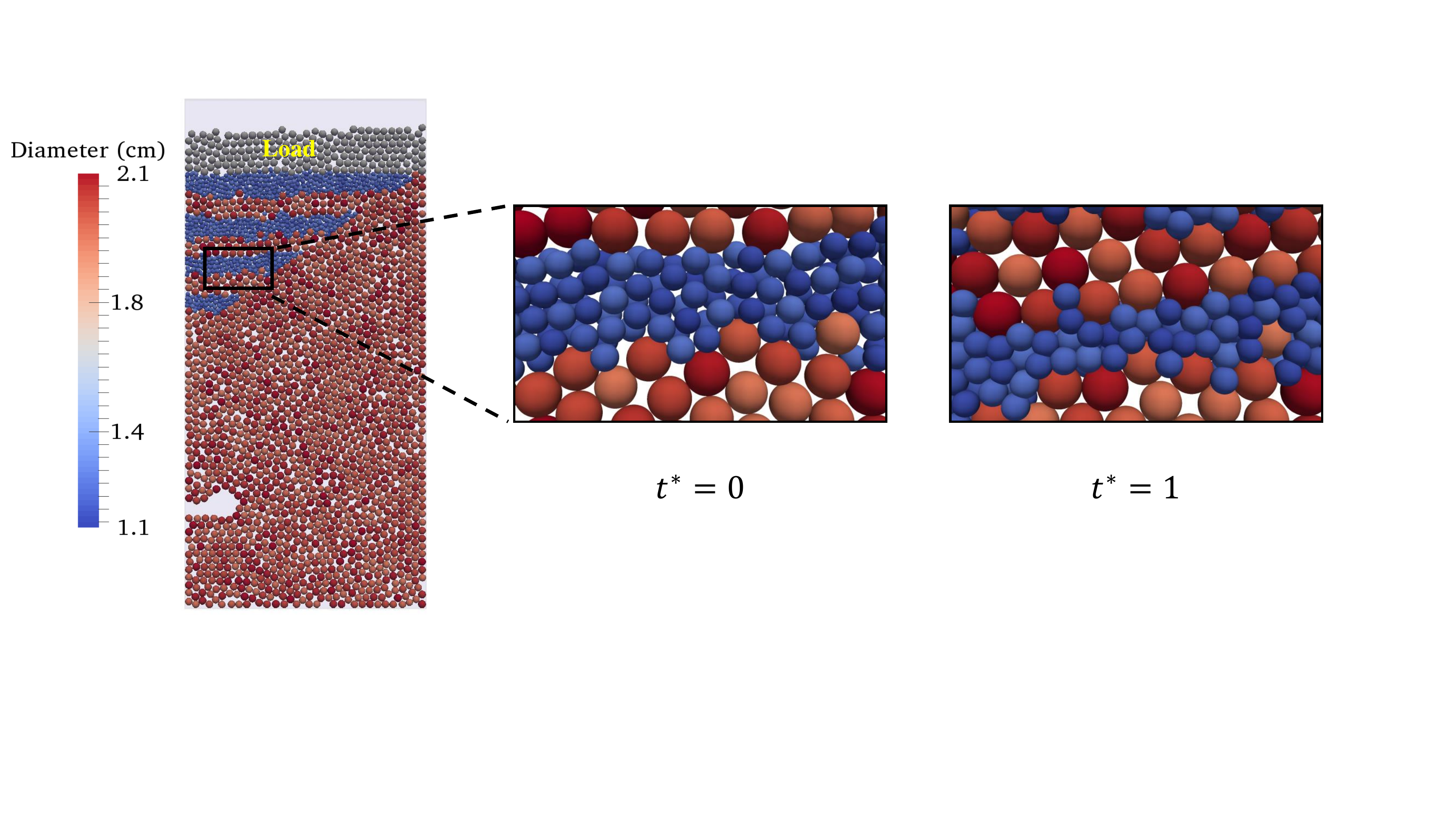}
    \caption{Initial and final structure of a layer of coke and pellet in a closer view.}
    \label{f:Softening-0-1}
\end{figure}

\begin{figure}[!tbp]
  \centering
    \includegraphics[trim = 20mm 0mm 0mm 0mm, clip, angle=0, width=17cm]{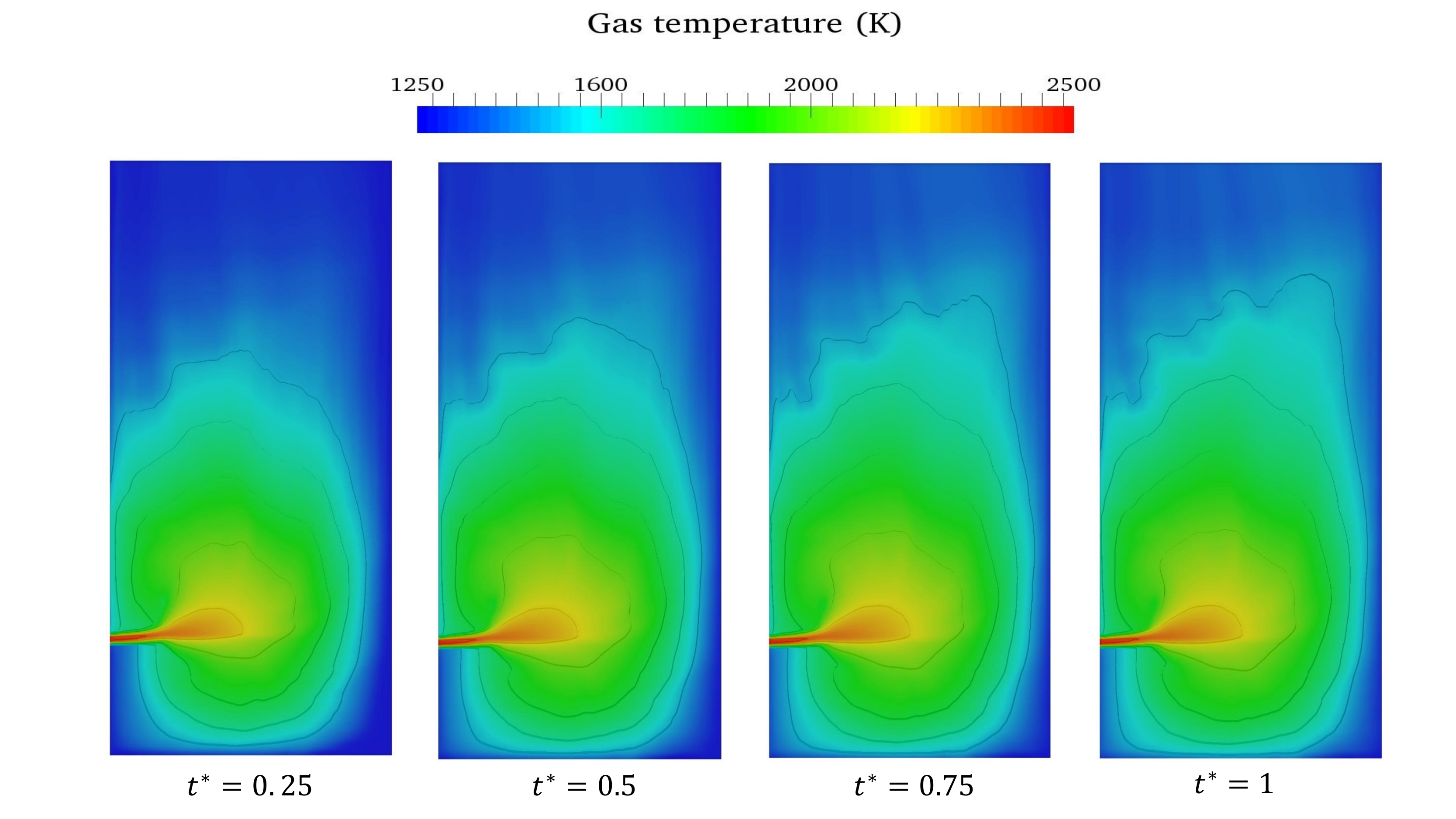}
    \caption{Gas temperature evaluation over time.}
    \label{f:gas_temperature}
\end{figure}

\begin{table}[ht]
 \caption{Thermal and mechanical properties of the reduced pellet and coke used in the XDEM simulation.}
 \resizebox{0.8\textwidth}{!}{
 \centering 
 \begin{tabular}{l l l}
  \hline
   Property&  Pellet& Coke  \\ \hline
      Thermal conductivity$(W m^{-1}K^{-1})$   & $0.5$ \cite{Takegoshi} & $0.96+0.00183*T$ \cite{coke} \\ [1ex]
      Heat capacity $(J kg^{-1}K^{-1})$  & $1000$  \cite{Takegoshi} & $320+0.61*T$ \cite{coke}  \\ [1ex]
      Intrinsic density$(kg m^{-3})$  & $3000$  &   $1000$ \\ [1ex]
      Coulomb friction coefficient & $0.7$  \cite{Gustafsson} &  $0.9$ \cite{Flo} \\ [1ex]
      Restitution factor  & $0.15$\cite{ASADA} & $0.2$   \cite{Flo}  \\ [1ex]
      Poisson ratio  & $0.24$ \cite{ASADA} & $0.3$ \cite{Flo} \\ [1ex]
  \hline
 \end{tabular}}
\label{table:properties}
\end{table}

% However the cohesive zone formation is strongly coupled with the reductions and the softening and melting behaviour of ore particles. 
% Therefore, without thermal and chemical reactions calculations, the cohesive zone has to be preset, including its shape and size.
The particles are heated up by absorbing convective heat from the hot gas passing through the void spaces. 
As the pellets are heated up, they start softening over the temperature due to the decrease of Young's modulus. 
Naturally, the lower pellet layer softens first due to facing the gas inlet earlier.
The height of the moving bed decreases more remarkably with further softening in the upper layers.
Figure~\ref{f:particle_temperature} shows the evaluation of particles temperature for different instances.
It can be seen
that the structural changes in the packed bed when the temperature increases and the position of the particles vary with time because the layers of pellets undergo shrinkage by particles' overlapping. 
%The decrease of the iron ore materials layers reach almost $40-50\%$ of the original height in the softening stage which is according to the real BF \cite{Bakker, Adema}. 
This phenomenon is presented in Figure~\ref{f:Softening-0-1} for a part of a layer in a closed view. It can be seen that the structure of the ore layer changes significantly from initial 
state $(t^*=0)$ to the final state $(t^*=1)$, however, the coke particles are not softened and remain intact due to their mechanical strength.
Additionally, the thermal energy of the gas is reduced as it travels through the packed bed due to the cooling effect of the particles.  
Thus, further upstream the gas temperature is reduced. 
On the other hand, as the bottom cokes and lower pellet layer are heated up and the temperature difference between them and the gas phase decreases, the hot gas can 
supply more thermal energy for the upper layers.
The evaluation of the gas temperature with iso-surfaces is shown in Figure~\ref{f:gas_temperature}.
One can observe how the heat absorption of coke and pellet induces 
changes in the gas temperature distribution, allowing clear identification of the cohesive zone by simply looking 
at the iso-temperature contours.

\begin{figure}[!tbp]
  \centering
    \includegraphics[trim = 0mm 0mm 0mm 0mm, clip, angle=0, width=18cm]{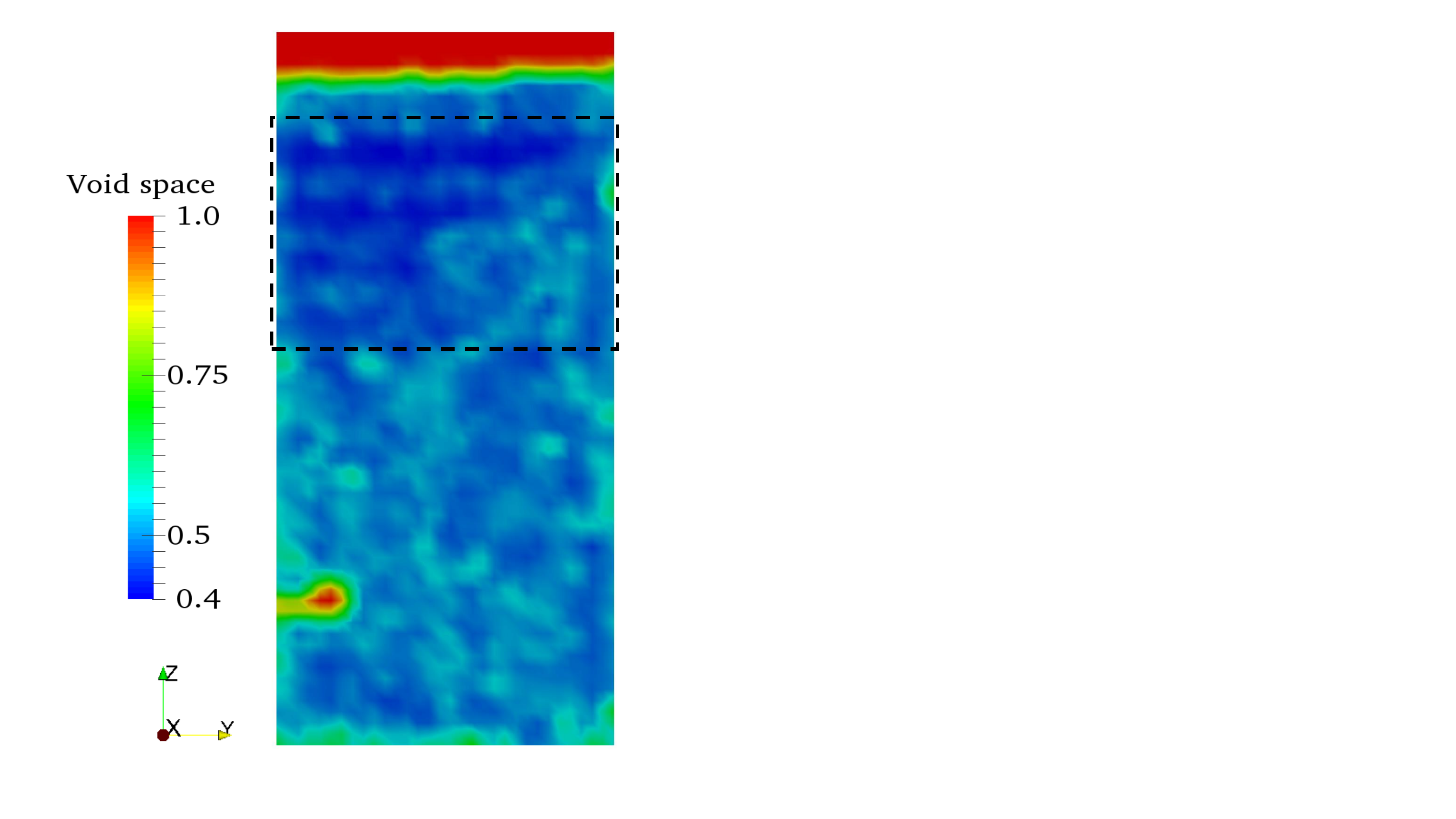}
    \caption{Void space distribution of the model geometry calculated by the XDEM.}
    \label{f:voidSpace}
\end{figure}

\begin{figure}[!tbp]
  \centering
    \includegraphics[trim = 80mm 0mm 0mm 0mm, clip, angle=0, width=18cm]{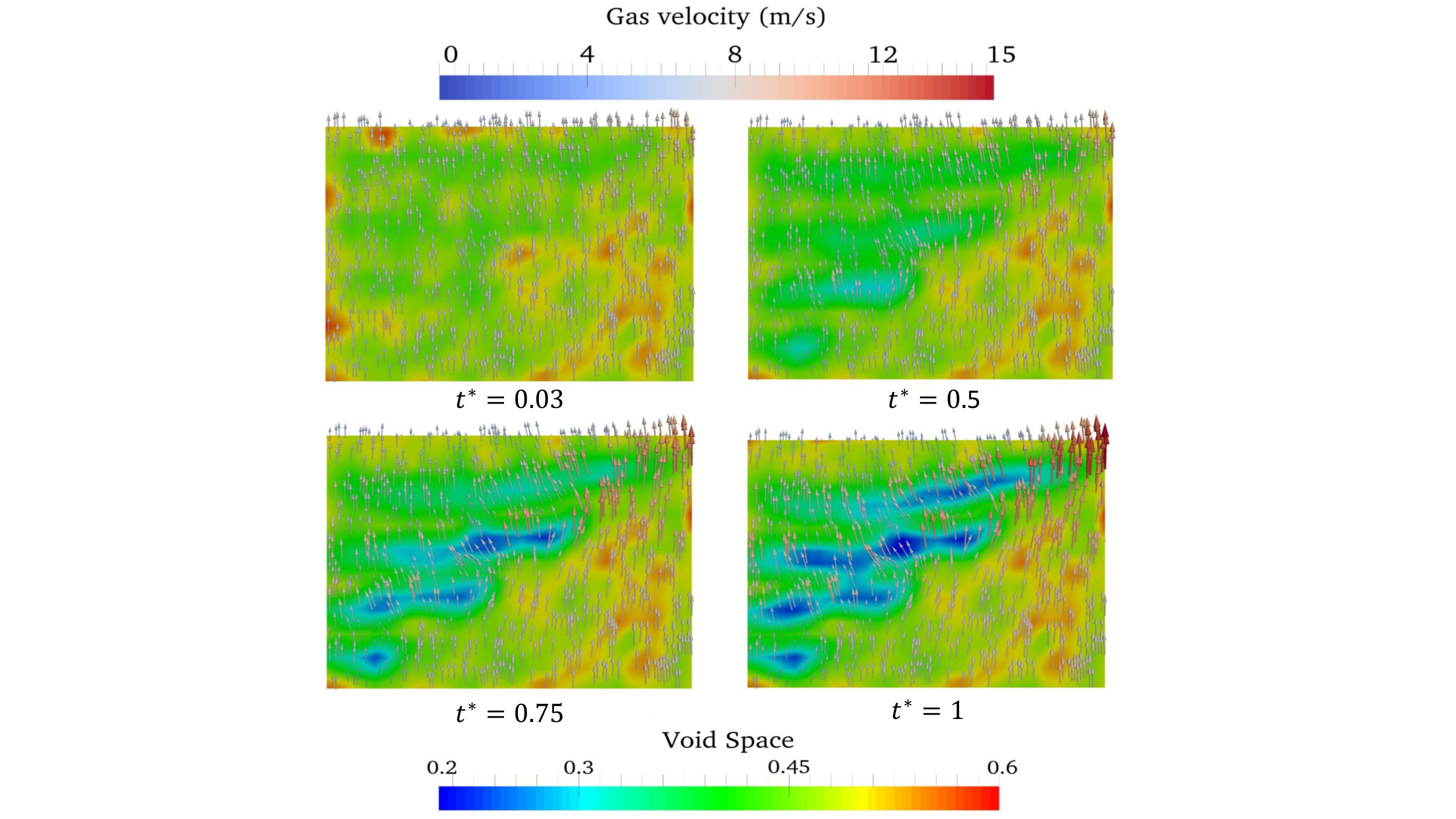}
    \caption{Gas vectors and void space evolution over time on the zone shown in Figure~\ref{f:voidSpace}.}
    \label{f:porosity-velocity}
\end{figure}

\begin{figure}[!tbp]
  \centering
    \includegraphics[trim = 0mm 0mm 0mm 0mm, clip, angle=0, width=15cm]{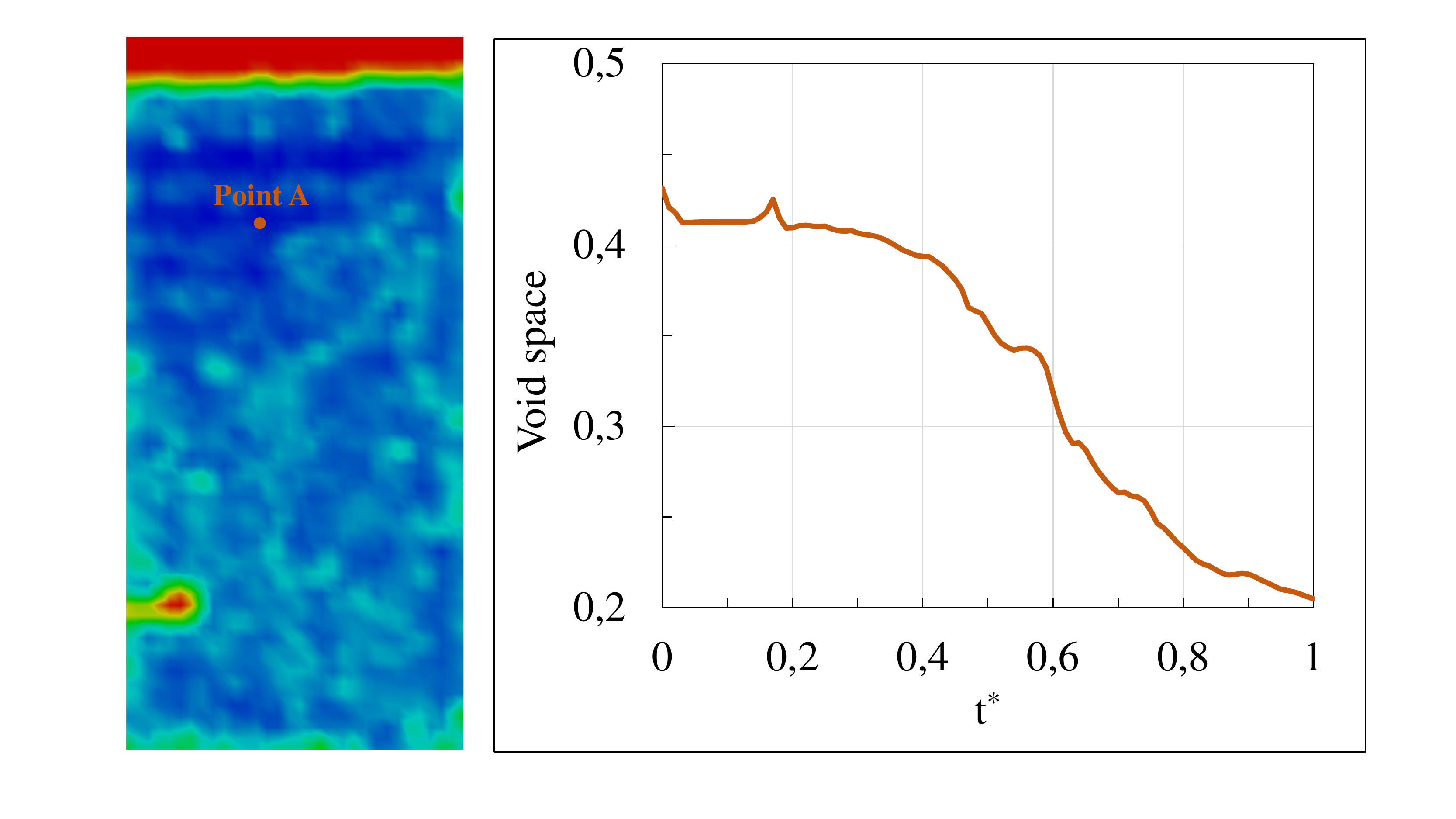}
    \caption{The local changes of the void space over time in the point A.}
    \label{f:localporo}
\end{figure}

As observed, the softening changes the local structures of the moving bed, void space between particles, and hence the gas flow. 
The initial void space distribution of the model geometry calculated by the XDEM is shown in Figure~\ref{f:voidSpace}. It is nearly $0.4$ for the iron ore pellets layers, $0.55$ for the coke layers, 
and $1$ in where no particles exist such as the raceway. 
To see the changes during the softening process in a better visualization, the evolution of void space and the gas flow is shown in Figure~\ref{f:porosity-velocity} for the marked region of 
Figure~\ref{f:voidSpace},
in where ore layers exist.
The void space is displayed in a contour plot, with vectors showing the gas velocity field.
It can be seen that, at the beginning, gas flows evenly upwards.
When the softening process begins, the overlap of pellet particles evolves, and subsequently, the void space of the pellet layers reduces and eventually reaches almost $0.2$. 
To better observe the phenomenon, the point A is selected and the local porosity changes over time for that point is presented in Figure~\ref{f:localporo}.
In this figure, it is possible to observe how the shrinking process starts at $t^*=0.2$, and up to
 $t^*=1$ the void space decreased to almost half of its initial value. It is also possible to observe how the process is 
 faster between $t^*=0.4$ and $t^*=0.8$ when the derivative of the void space over time is clearly changing.
 This phenomenon was already observed, under simpler conditions, in Figure \ref{f:exp_simulation}, and confirms
 once more the non-linear nature of the bed-shrinkage.  
% On another hand, the void space of coke layers remain at the initial value. 
 
\begin{figure}[!tbp]
  \centering
    \includegraphics[trim = 0mm 0mm 0mm 0mm, clip, angle=0, width=15cm]{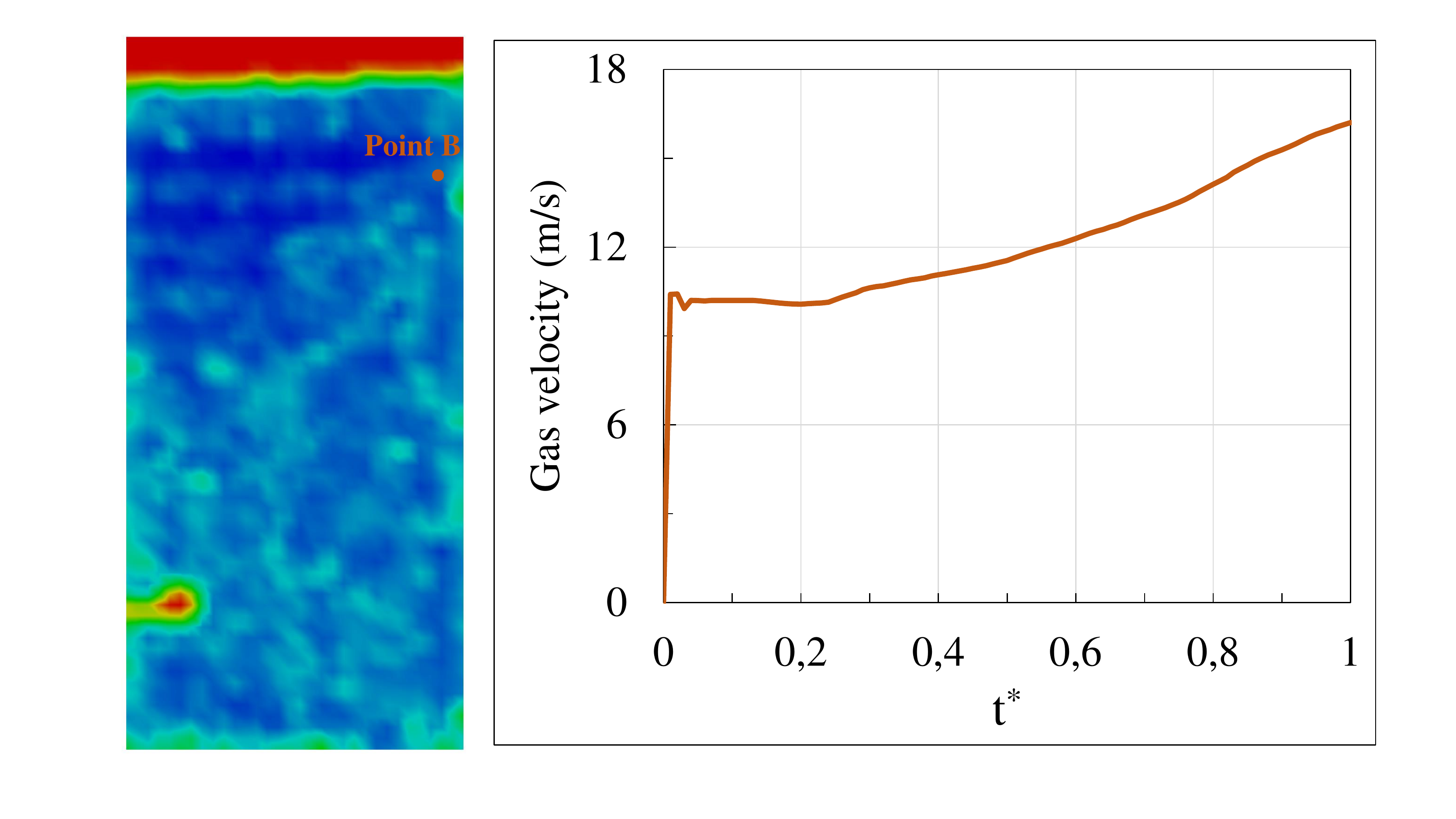}
    \caption{The local changes of the gas velocity over time in the point B.}
    \label{f:localvelo}
\end{figure}

\begin{figure}[!tbp]
  \centering
    \includegraphics[trim = 20mm 0mm 0mm 0mm, clip, angle=0, width=17cm]{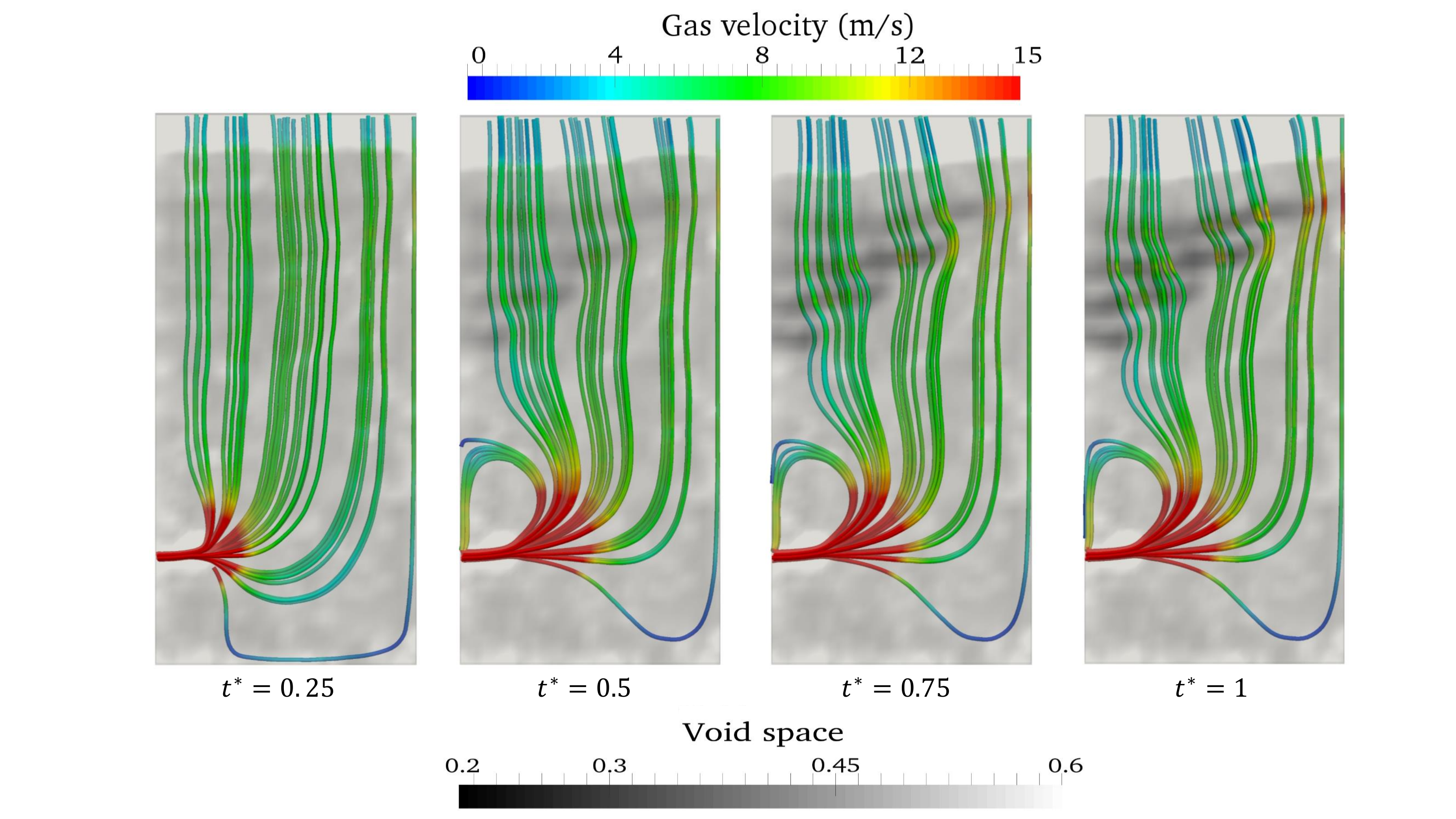}
    \caption{Gas stream lines over time.}
    \label{f:StreamLines}
\end{figure}

\begin{figure}[!tbp]
  \centering
    \includegraphics[trim = 5mm 0mm 0mm 0mm, clip, angle=0, width=15cm]{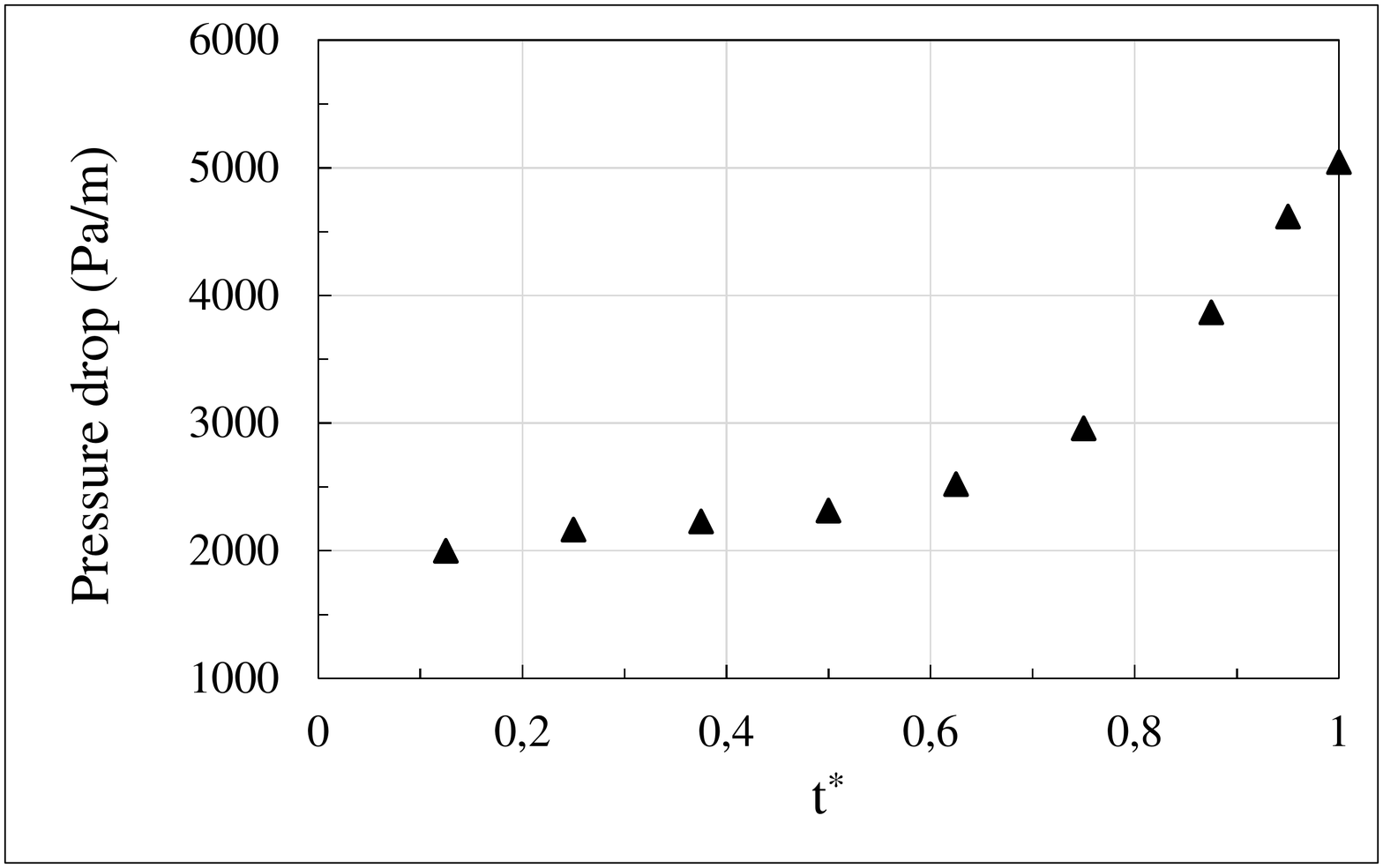}
    \caption{Pressure drop changes over time on the zone shown in Figure~\ref{f:voidSpace}.}
    \label{f:deltaP}
\end{figure}

The reduction of the void space affects the gas flow significantly in both magnitude and direction.
In particular, as shown in Figure~\ref{f:porosity-velocity}, the gas flow loses its uniformity. The velocity vectors can 
be observed to bend in the proximity of the low-porosity zones and find alternative paths toward higher porosity regions.
Figure~\ref{f:localvelo} shows the evaluation of the gas velocity over time at the center of the bed for the  point B. 
It is interesting to observe how the velocity  magnitude in point B is clearly correlated to the 
porosity changes shown in Figure~\ref{f:porosity-velocity}. In particular, the rise in velocity starts around $t^*=0.2$, and reaches almost $1.5$ times 
its original value at $t^*=1$. This underlines the importance of the mutual interaction between the gas flow and the 
particles dynamics and therefore the significance of a numerical method able to resolve them together.
In order to track the gas flow pattern in the whole domain from the entrance to the top, the stream lines of the gas flow colored by their magnitude are presented in Figure~\ref{f:StreamLines}. 
It should be noted that the gas 
flow is not the only hydrodynamic parameter affected by the softening process. This phenomenon has also a profound effect on the pressure drop.
This basic fluid-dynamic relation is here particularly delicate. As a matter of fact the pressure drop in static conditions
is known to be a function of both porosity and velocity. Due to the fact that within the CZ of the BF the 
porosity changes dynamically, and, as a consequence, it changes dynamically  the gas velocity, pressure drop predictions 
based on static conditions are no longer accurate. This is a further motivation to apply CFD-DEM models to study the 
BF process, as the extremely important operative parameter of the pressure drop is very difficult to be predicted else way.
In Figure~\ref{f:deltaP} the obtained pressure drop over the height of the specified zone is depicted. As can be seen, the pressure drop increased over time due to the development of the softening process.
The pressure drop is enhanced with a sharper slope due to the low value of void space when the particle's temperatures are
close to the melting point.

As it observed the proposed approach has the possibility of monitoring
 global features of the flow with local parameters in the CZ, and opens the path for the possible 
 future developments to study the CZ of the BF in a real condition.

% 
% \begin{figure}[!tbp]
%   \centering
%     \includegraphics[trim = 0mm 0mm 0mm 0mm, clip, angle=0, width=18cm]{Figures/NewResolution_old/gas_vectors_new.pdf}
%     \caption{Gas vectors and void space evolution over time on the zone shown in Figure~\ref{f:voidSpace}.}
%     \label{f:porosity-velocity}
% \end{figure}

\section{Conclusions}
The eXtended Discrete Element Method (XDEM) has been applied to describe the softening behavior of pellets in 
the cohesive zone of an experimental blast furnace. The XDEM is a CFD-DEM method coupled with heat transfer between gas-particle and particle-particle conduction and
radiation that are significant for the softening process as well as for all processes taking place in the BF. The XDEM is also enhanced with dual-grid multiscale approach coupling which is a necessity when 
the fluid velocity is high, in particular, in the raceway of a blast furnace. In conclusion, the developed model is verified for gas-solid behaviour during the softening process, in which the deformation 
of particles is expressed by the overlapping approach. In this study, the parameters such as the deformation, displacement, and temperature of the particles on the one hand and the gas phase temperature, velocity, 
and pressure drop, on the other hand, were analyzed. In the end, the mutual effects of the gas and solids were comprehensively studied. Although, any reactions and melting process
were not considered, 
it was observed that the findings are useful to establish a model for identification of a cohesive zone of a blast furnace in the real condition.
\newline

\textbf{Acknowledgements}\newline
The authors are thankful to the Luxembourg National Research Fund (FNR) and the HPC group of the University of Luxembourg.

\begin{table}%
 \resizebox{0.6\textwidth}{!}{
\begin{tabular}{l l }%
\multicolumn{2}{l}{\textbf{ Nomenclature}}\\

$g$&gravitational accelerating $(m s^{-2})$\\

$t$&time $(s)$\\

$T$&temperature $(K)$\\

%
%$U$&velocity,  {\metre\per\second}\\
$h$&enthalpy $(J kg^{-1})$\\

$r$&particle radius $(m)$\\
$m$&particle mass $(kg)$\\
$x$&position $(m)$\\
$q'$&heat flux $(W m^{-2})$\\

$v$&velocity $(m s^{-1})$\\

$V$&volume $(m^{3})$\\

$p$&pressure $(Pa)$\\

$F$&force $(N)$\\

$I$&momentum of inertia $(j kg^{-1})$\\

$M$&torque $(N m)$\\

%%%%%%%%%%%%%%%%%%%%%%%%%%%%%%%%%%%%%%%%%%%%%%%%%%%%%%%%%%%%%%%%%%%%%
%
%
%%%%%%%%%%%%%%%%%%%%%%%%%%%%%%%%%%%%%%%%%%%%%%%%%%%%%%%%%%%%%%%%%%%%%

\multicolumn{2}{l}{{\textbf{Greek Symbols}}}\\
%%%%%%%%%%%%%%%%%%%%%%%%%%%%%%%%%%%%%%%%%%%%%%%%%%%%%%%%%%%%%%%%%%%%%
%%%%%%%%%%%%%%%%%%%%%%%%%%%%%%%%%%%%%%%%%%%%%%%%%%%%%%%%%%%%%%%%%%%%%
$\epsilon$&void space $(-)$\\

$\alpha$&heat transfer coefficient $(W m^{-2} K^{-1})$\\
$\lambda$&thermal conductivity  $(W m^{-1} K^{-1})$\\
%
%$\espilon_i$&phase volume fraction, -\\

$\rho$&density $(kg m^{-3})$\\
%\\
%$\beta$&mass transfer coefficient,  {\metre\per\second}\\
%\\
%$\gamma$&-&interpolation coefficient\\
%$\Gamma$&-&diffusion coefficient\\
%\\
%$\epsilon_P$&porosity within a porous particle, -\\
%

%
%
$\mu$&dynamic viscosity $(kg m^{-1} s^{-1})$\\
$\delta$&overlap $(m)$\\

$\omega$&angular velocity $(rad s^{-1})$\\

\multicolumn{2}{l}{\textbf{Superscripts and Subscripts}}\\
$i, j$&particle\\
$p$&particle\\
$c$&contact forces\\
$n$&normal direction\\
$t$&tangential direction\\
$rad$&radiation\\
$cond$&conduction\\
$inf$&ambient\\
$*$&dimensionless\\
$g$&gas\\
$s$&solid\\
$d$&drag\\

\end{tabular}}
\end{table}

\textbf{References}\newline

  \bibliographystyle{elsarticle-num}





% \bibitem{Feynman1963118}"iuwdhie"
% 
% \end{thebibliography}
\end{document}